\documentclass[aps, prd, twocolumn, nofootinbib, floatfix, 
superscriptaddress]{revtex4-2}
\pdfoutput=1

\usepackage{subfigure}
\usepackage{epsfig}
\usepackage{amsmath}
\usepackage{amsfonts}
\usepackage{amssymb}
\usepackage{amssymb,amsmath,amsfonts}
\usepackage{xfrac}
\usepackage{color}
\usepackage[utf8]{inputenc}
\usepackage{graphicx}
\usepackage{dcolumn}
\usepackage{bm}
\usepackage{tikz}

\usepackage{tcolorbox}
\usepackage{hyperref}
\hypersetup{colorlinks, citecolor=red, linkcolor=bluscuro, urlcolor=bluscuro}
\definecolor{rossos}{cmyk}{0,1,1,0.55}
\definecolor{bluscuro}{rgb}{0.15, 0.2, .85}
\definecolor{bluchiaro}{cmyk}{1,.3,0.,0.1}

\begin{document}

\title{Topological dark energy from black-hole formations and mergers through 
the gravity-thermodynamics approach}

\author{Stylianos A. Tsilioukas}
\email{tsilioukas@sch.gr}
\affiliation{\mbox{Department of Physics, University of Thessaly, 35100 Lamia, 
Greece}}
\affiliation{\mbox{National Observatory of Athens, Lofos Nymfon, 11852 Athens, 
Greece}}
 
 \author{Nicholas Petropoulos}
 \email{npetropoulos@uth.gr}
 \affiliation{\mbox{Department of Physics, University of Thessaly, 35100 Lamia, 
Greece}}

\author{Emmanuel N. Saridakis}
\email{msaridak@noa.gr}
\affiliation{\mbox{National Observatory of Athens, Lofos Nymfon, 11852 Athens, 
Greece}}
\affiliation{CAS Key Laboratory for Researches in Galaxies and Cosmology, 
Department of Astronomy, \\
University of Science and Technology of China, Hefei, 
Anhui 230026, P.R. China}
\affiliation{\mbox{Departamento de Matem\'{a}ticas, Universidad Cat\'{o}lica 
del 
Norte, 
Avda.
Angamos 0610, Casilla 1280 Antofagasta, Chile}}

\begin{abstract}

We apply the  gravity-thermodynamics approach in the case of 
Einstein-Gauss-Bonnet theory,  and its corresponding Wald-Gauss-Bonnet entropy, 
which due to the Chern-Gauss-Bonnet theorem it is related to the Euler 
characteristic of the Universe topology. However, we consider  the realistic 
scenario where   we have the formation and  merger of black holes 
that lead to  topology changes, which  induce entropy changes in the Universe 
horizon. We  extract the modified Friedmann equations  
and we obtain an effective  dark energy sector of topological origin. We 
estimate the black-hole formation and merger rates starting from the observed  
star formation rate per redshift, which is parametrized very efficiently by the 
Madau-Dickinson form, and finally we obtain a dark-energy energy density 
that depends   on the   cosmic star formation rate density, on the
fraction $f_{\text{BH}}$ of stars forming black holes, on the fraction of black 
holes $f_\text{merge}$ that eventually merge, on the  fraction  $ 
f_{\text{bin}}$ of massive stars that are in binaries, on the   average mass of 
progenitor stars that will evolve to form black holes \( \langle 
m_{\text{prog}} 
\rangle \), as well as on the Gauss-Bonnet coupling constant.
We investigate in detail the cosmological evolution, obtaining the  usual 
thermal history.  Concerning  the dark-energy equation-of-state parameter, we 
show that at intermediate redshifts it exhibits phantom-like and 
quintessence-like behavior according to the sign of the  Gauss-Bonnet coupling, 
while at   early and late times it tends to the cosmological constant value.
 Finally, we study the effect of the other model parameters,  showing that  
for  the  whole allowed observationally estimated ranges, the topological 
dark-energy equation-of-state parameter remains within its observational bounds.

\end{abstract}

\maketitle

\section{Introduction}\label{Introduction}

According to extensive observational evidence from various origins, the 
Universe 
has recently entered a phase of accelerated expansion \cite{Riess:1998cb, 
Perlmutter:1998np, Spergel:2003cb, Tegmark:2003ud, Allen:2004cd, 
Astier:2005qq}. 
In order to explain this behavior two main directions have been pursued. The 
first is to retain general relativity as the gravitational framework while 
introducing new energy components, such as the dark energy sector 
\cite{Copeland:2006wr, Cai:2009zp}. The second involves constructing modified 
and extended theories of gravity by altering the left-hand side of Einstein 
field equations, adding correction terms to the standard Einstein-Hilbert 
action \cite{CANTATA:2021asi,Capozziello:2011et,Cai:2015emx,Nojiri:2017ncd}. 
Such modified gravity theories not only address cosmological issues but also 
offer improved quantum behavior \cite{AlvesBatista:2023wqm}, given that general 
relativity is non-renormalizable \cite{Addazi:2021xuf}. Specifically, 
incorporating higher-order curvature terms into the Einstein-Hilbert Lagrangian 
tends to eliminate divergences \cite{Stelle:1976gc}, and among all   
higher-order terms the Gauss-Bonnet (GB) combination is special since it is 
topologically invariant in 4 dimensions and thus plays a significant role in 
heterotic string theory 
\cite{Stelle:1976gc,Zwiebach:1985uq,Boulware:1985wk,Gross:1986mw} and in 
M-theory \cite{Vafa:1994tf}. 

Beyond the aforementioned approaches to modified gravity theories, there exists 
a well-known conjecture that gravity can be described through the laws of 
thermodynamics \cite{Jacobson:1995ab,Padmanabhan:2003gd,Padmanabhan:2009vy}. 
This concept is inspired from black-hole thermodynamics, where a black hole 
(BH) 
is assigned a specific temperature and entropy, dependent on its horizon 
\cite{Gibbons:1977mu}. The  ``thermodynamics of spacetime" conjecture 
\cite{Jacobson:1995ab} draws an analogy from BH horizon thermodynamics to the 
Universe horizon at  cosmological scales. In particular, if one applies the 
first law of thermodynamics in the apparent horizon one can obtain the 
Friedmann equations \cite{Cai:2005ra,Akbar:2006er,Cai:2006rs}, and vice versa 
the Friedmann equations can be expressed as the first law of thermodynamics. 
This procedure has been applied successfully both in the case of general 
relativity and as well as in modified theories of gravity 
\cite{Paranjape:2006ca, Sheykhi:2007gi, Akbar:2006kj, Jamil:2009eb, Cai:2009qf, 
Wang:2009zv, Jamil:2010di, Gim:2014nba, Fan:2015aia, 
Lymperis:2018iuz,Saridakis:2020zol,Basilakos:2023kvk}, where in the latter case 
 
 one should use the corresponding modified entropy relation.

In the present work we are interested  in applying the gravity-thermodynamics 
approach in the case where the gravitational action   is extended by the GB 
term and the corresponding entropy is extended by the Wald-Gauss-Bonnet 
entropy \cite{Wald:1993nt,Iyer:1994ys}, a term dependent on the BH horizon 
topology. However, we are interested in considering the realistic scenario 
where in the Universe we have the formation and  merger of black holes, that 
lead to  topology changes which  induce entropy 
changes in the Universe horizon. Hence, applying the
gravity-thermodynamics analysis one could  extract  modified Friedmann 
equations,  with  an effective, dark  energy sector of topological origin, 
depending on the black-hole formation and merger rates.

The plan of this manuscript is the following: In section \ref{the model} we 
review the standard gravity-thermodynamics approach, and then we apply it in 
the case of Einstein-Gauss-Bonnet theory, extracting the 
modified Friedmann equations. In section \ref{Cosmic evolution} we first 
provide an estimation for the   BH formation and merging rates using the star 
formation rate as the starting point, and then we proceed to the detailed 
cosmological applications of specific scenarios, focusing on the behavior of 
the dark-energy and matter density parameters, of the effective  dark-energy 
equation-of-state parameter, and of the deceleration parameter. 
  Finally in \ref{Conclusions} we summarize our results and we conclude.

\section{Modified cosmology from Wald-Gauss-Bonnet entropy}
\label{the model}

In this section we present the scenario at hand, extracting the modified 
Friedmann 
equations by applying the first law of thermodynamics to the Universe using the 
Wald-Gauss-Bonnet entropy. We start by considering a homogeneous and isotropic 
Friedmann-Robertson-Walker (FRW) geometry with metric
\begin{equation}
\mathrm{d}s^2 = -\mathrm{d}t^2 + a^2(t) \left( \frac{\mathrm{d}r^2}{1 - k r^2} 
+ 
r^2 \mathrm{d}\Omega^2 \right),
\end{equation}
where \(a(t)\) is the scale factor, and \(k = 0, +1, -1\) corresponds to flat, 
closed, and open spatial geometry, respectively.

\subsection{Friedmann equations as the first law of 
thermodynamics}\label{Friedmann equations}

To extract the Friedmann equations in general relativity from the first law of 
thermodynamics, we consider an expanding Universe filled with a matter perfect 
fluid, with energy density \(\rho_m\) and pressure \(p_m\). As a boundary, ones 
uses the apparent horizon \cite{Cai:2005ra,Akbar:2006er,Cai:2006rs}
\begin{equation}\label{apparent horizon radius}
\tilde{r}_A = \frac{1}{\sqrt{H^2 + \frac{k}{a^2}}},
\end{equation}
where \(H = \frac{\dot{a}}{a}\) is the Hubble parameter. The apparent horizon 
is a marginally trapped surface with vanishing expansion, and it is a causal 
horizon associated with gravitational entropy and surface gravity 
\cite{Jacobson:1995ab,Padmanabhan:2003gd,Padmanabhan:2009vy}.
The temperature attributed to the apparent horizon is
\begin{equation}\label{apparent horizon temperature}
T_h = \frac{1}{2\pi \tilde{r}_A},
\end{equation}
and the entropy in general relativity is given by the standard 
Bekenstein-Hawking relation
\begin{equation}\label{Bekentein Hawking entropy}
S_h = \frac{A}{4G},
\end{equation}
where \(A = 4\pi \tilde{r}_A^2\) is the area of the apparent horizon. Assuming 
that the Universe fluid acquires the same temperature with the horizon after 
equilibrium, the heat flow crossing the horizon during an infinitesimal time 
interval \(dt\) can be found to be \cite{Cai:2005ra}
\begin{equation}\label{dE 1st law}
\delta Q = -\mathrm{d}E = A (\rho_m + p_m) H \tilde{r}_A \mathrm{d}t.
\end{equation}
Since the first law of thermodynamics states that \(-\mathrm{d}E = T 
\mathrm{d}S\), 
using the temperature and entropy from \eqref{apparent horizon temperature} and 
\eqref{Bekentein Hawking entropy}  we find
\begin{equation}\label{2nd Friedmann LCDM}
-4 \pi G (\rho_m + p_m) = \dot{H} - \frac{k}{a^2}.
\end{equation}
Finally, considering that the matter fluid satisfies the conservation equation
\begin{equation}\label{conservation equation}
\dot{\rho}_m + 3H(\rho_m + p_m) = 0,
\end{equation}
 integrating \eqref{2nd Friedmann LCDM} we obtain
\begin{equation}\label{1_FriedmannLCDM}
\frac{8 \pi G \rho_m}{3} = H^2 + \frac{k}{a^2} - \frac{\Lambda}{3},
\end{equation}
with \(\Lambda\) the integration constant. Equations \eqref{1_FriedmannLCDM} 
and  \eqref{2nd Friedmann LCDM} are nothing else but the two Friedmann 
equations, with the integration constant \(\Lambda\) 
playing the role of the cosmological 
constant. 

In summary, through the gravity-thermodynamics approach one can 
extract the Friedmann equations not by varying the gravitational action, but 
by applying the first law of thermodynamics on the Universe apparent horizon. 
Hence, this implies that one can obtain modified  Friedmann equations  without 
necessarily  needing to alter the underlying gravitational theory, but just 
altering the entropy expression itself.

\subsection{Wald-Gauss-Bonnet entropy}\label{Wald Gauss Bonnet entropy}

In the previous subsection we saw how one can extract the Friedmann equations 
through the gravity-thermodynamics conjecture. As we mentioned in the 
Introduction, the same procedure  can be applied  by changing the entropy 
relation.

One of the widely studied higher-order theories of gravity is the Gauss-Bonnet 
one, due 
its topological significance \cite{Lanczos:1938sf,Lovelock:1971yv}. The total 
action in four dimensions consists of the Einstein-Hilbert and the Gauss-Bonnet 
(GB) term, namely 
\begin{equation}\label{EH GB action}
    I = \frac{1}{16 \pi G} \int d^4 x \sqrt{-g} \left( R + \tilde{\alpha} 
\mathcal{G} \right),
\end{equation}
where $R$ is the Ricci scalar,
$\mathcal{G}$ is the GB term, defined as
\begin{equation}
    \mathcal{G} = R^2 - 4 R_{\mu\nu} R^{\mu\nu} + R_{\mu\nu\rho\sigma}
R^{\mu\nu\rho\sigma},
\end{equation}
and
 $\tilde{\alpha}$ is the GB coupling constant. Note that in four 
dimensions the  Gauss-Bonnet term that enters linearly in the action does not 
lead to any new term in the 
field equations, since it is a topological invariant, i.e. the above 
Einstein-Gauss-Bonnet action   is just General Relativity. However, this term 
does lead to new contributions in the entropy expression.
    
    Let us calculate the corrections to the entropy expression.
The Wald-Noether charge method \cite{Wald:1993nt,Iyer:1994ys} can 
be used to derive a horizon entropy $S$ satisfying the first law of 
thermodynamics for any first-order spacetime perturbation. For any 
diffeomorphism invariant theory of gravity described by a Lagrangian \(L\), the 
Wald entropy of a stationary BH with a regular bifurcation surface is given by 
\cite{Iyer:1994ys,Jacobson:1993xs}
\begin{equation}
    S_{\text{Wald}} = -\frac{1}{8} \int_B \frac{\partial L}{\partial R_{ijkl}} 
\epsilon_{ij} \epsilon_{kl} \sqrt{\sigma} \, d^{D-2}x,
\end{equation}
where the integration is over any \((D-2)\)-dimensional spacelike cross-section 
\(B\) of the horizon, and \(\epsilon_{ij}\) is the binormal on such a 
cross-section, normalized as \(\epsilon_{ij} \epsilon^{ij} = -2\). For the 
Einstein-Gauss-Bonnet action   \eqref{EH GB action}, the Wald 
entropy becomes
\begin{equation}
    S_{\text{WGB}} = \frac{1}{4} \int_B \left(1 + 2\tilde{\alpha} {}^{(D-2)}R 
\right) \sqrt{\sigma} \, d^{D-2}x,
\end{equation}
where \({}^{(D-2)}R\) is the Ricci curvature associated with the 
\((D-2)\)-dimensional cross-section of the horizon. Applying   to $D=4$ 
dimensions  the expression of the Wald-Gauss-Bonnet (WGB) entropy becomes
\begin{equation}
    S_{\text{WGB}} = \frac{A}{4G} + \frac{\tilde{\alpha}}{2G} \int_{h} d^2 x 
\sqrt{\sigma} \mathcal{R},
\label{SWGB1}
\end{equation}
where \(\mathcal{R}\) is the Ricci scalar of the induced metric on the horizon 
$h$.

Now, according to the Chern-Gauss-Bonnet theorem  \cite{Chern:1945}, the 
integral of the Ricci scalar \( \mathcal{R} \) over the two-dimensional horizon 
\( h \) equals \( 4\pi \) times the Euler characteristic \( \chi(h) \) of the 
horizon, namely
\begin{equation}
    \int_{h} d^2 x \sqrt{\sigma} \mathcal{R} = 4\pi \chi(h).
    \label{Eulerchar}
\end{equation}
The Euler characteristic $\chi$ is defined as the alternating sum of 
the 
Betti numbers   $B_{p}$ of a manifold $M$, namely $  
\chi(M)=\sum_{p}(-1)^{p}B_{p}$
\cite{Nakahara:2003nw},
where the Betti numbers  are defined as the dimension of 
the $p^{th}$ de Rahm cohomology group $H^{p}(M)$, i.e.
$    B_{p}=dim H^{p}(M)$ \cite{Tsilioukas:2023tdw,Tsilioukas:2024tjh}. 
Hence, combining (\ref{SWGB1}) and (\ref{Eulerchar}) we find that the entropy 
for a BH in the presence of the 
Einstein-Gauss-Bonnet terms is \cite{Sarkar:2010xp}
\begin{equation}\label{WGB entropy}
   S_{\text{WGB}} = \frac{A}{4G} + \frac{2\pi \tilde{\alpha}}{G} \chi(h). 
\end{equation}
Thus, the Wald-Gauss-Bonnet entropy can be interpreted as a sum of an 
area term (the Bekenstein-Hawking one) and a topological term (the Euler 
characteristic of the horizon), namely
\begin{equation}\label{SWGB terms}
    S_{\text{WGB}}=S_{\text{area}} + S_{\text{top}}.
\end{equation}

\begin{figure}[!htbp]
        \centering
        \includegraphics[width=0.38\textwidth]{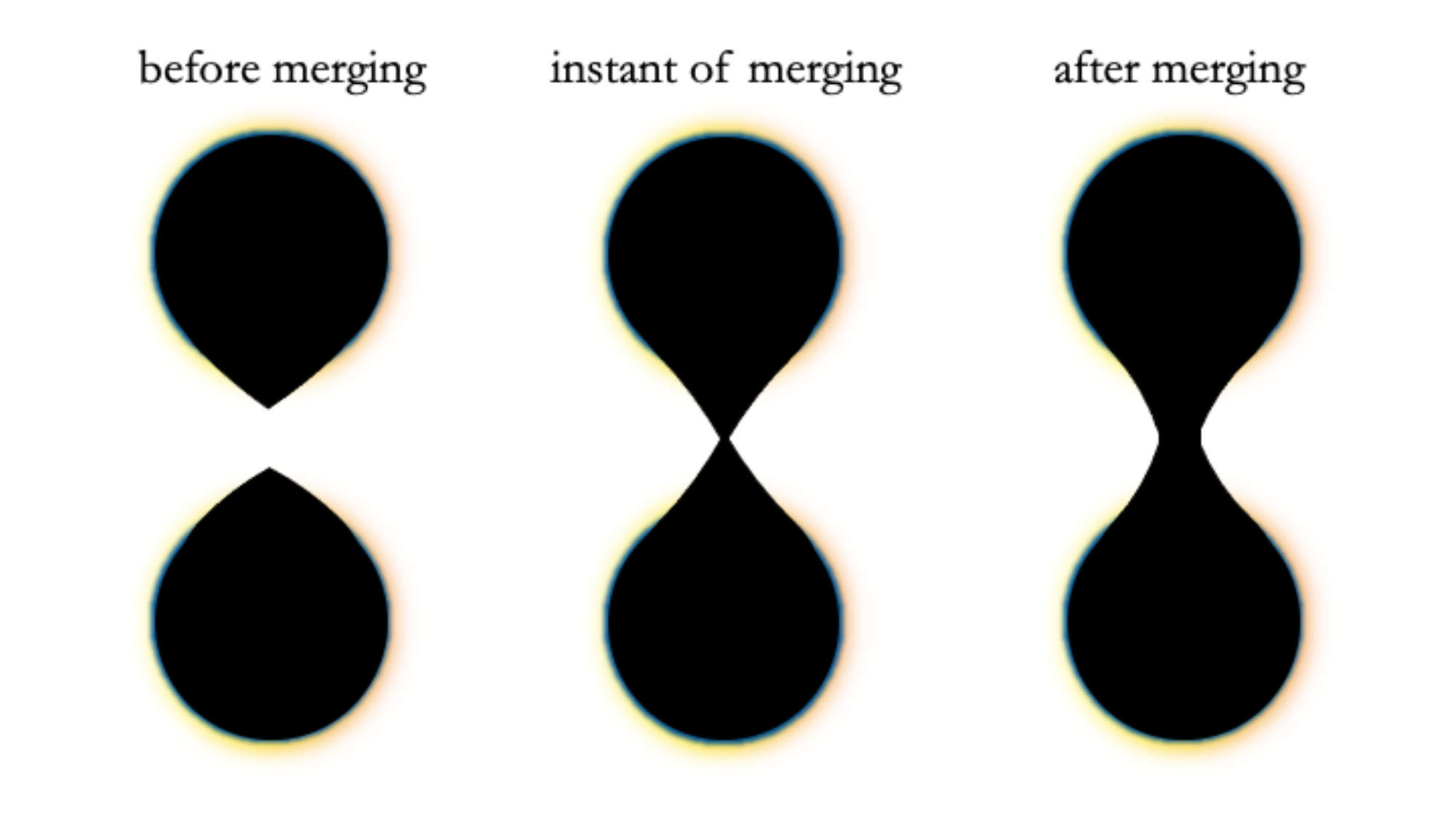}    
\includegraphics[width=0.38\textwidth]{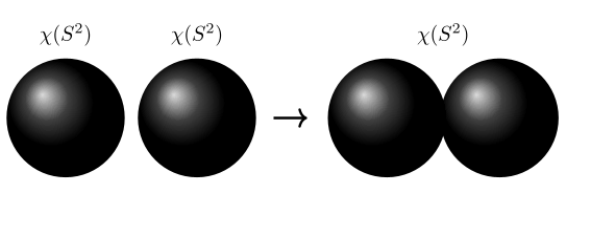}
    \caption{ \textit{The topology and the  Euler characteristic change 
of a 
black hole merger.} }
    \label{fig:BH merge}
\end{figure}

Nevertheless, the above consideration may lead to a potential problem when we 
have black hole mergers. As it was discussed in 
  \cite{Liko:2007vi,Sarkar:2010xp}, while 
in the case of general relativity the corresponding Bekenstein-Hawking entropy 
increases, the Wald entropy of the Einstein-Gauss-Bonnet theory 
exhibits a decrease of topological origin.  In particular, as we saw above,
in the case of the GB term the corresponding entropy \eqref{WGB entropy} 
depends on the Euler characteristic of the horizon $\chi(h)$. Since the 
horizon of a 
BH is a two-dimensional sphere $S^2$, during the merger of two BHs  their 
horizons   merge too, as illustrated in Fig.~\ref{fig:BH merge}.
Before the merger, the spacetime contains two horizons,  i.e.
\begin{equation}
    \chi_{in}=\chi(S^2) + \chi (S^2)=4,
\end{equation}
while  after the merger there is only one horizon, namely
\begin{equation}
    \chi_{f}=\chi(S^2)=2,
\end{equation}
 and therefore  during a black hole merger one obtains an Euler characteristic 
change
\begin{equation}\label{delta chi}
    \delta \chi_{h} = \chi_{f} - \chi_{in} = -2.
\end{equation}
Hence, if one considers  Einstein-Gauss-Bonnet theory and the extended entropy 
\eqref{WGB entropy} it is straightforward to see that  for $\tilde{\alpha}>0$, 
the topological transition  during the BH 
merger  induces a decrease in the topological part of Wald entropy $\Delta 
S_{top} \leq 0$, resulting to a violation of the second law (in the same 
lines, when a BH horizon forms from gravitational collapse  an increase of the 
Wald entropy is induced by the topological transition $\delta \chi = 2$).  On 
the other hand, in the case 
$\tilde{\alpha}<0$ there is an increase of Wald entropy during BH mergers and a 
decrease during a BH formation.

Since the above violation originates from the decrease of the BHs horizon Euler 
characteristic, and since the decrease is an instantaneous integer jump  
which 
cannot be compensated by a continuous procedure \cite{Liko:2007vi,Sarkar:2010xp}, one 
possible way to resolve the second law violation is to assume that a similar 
process exists that compensates the decrease of the topological part of the 
entropy. Having in mind the discussion of the gravity-thermodynamics conjecture 
and 
that the BH horizon and the apparent horizon are causal 
horizons, we assume here that the topology of the causally 
connected boundaries remains constant. In particular, the apparent horizon 
is the external 
boundary while the BH horizons are the internal ones, and therefore we can 
define the total boundary as $\partial M = \mathcal{H} \bigcup_{i =1 }^{N} 
h_{i}$ and then apply the calculation for the Euler characteristics  as 
\begin{equation}
    \chi (\partial M) = \chi (\mathcal{H}) + \sum_{i =1 }^{N} \chi( h_{i}).
\end{equation}
Since the overall boundary topology remains constant $\delta \chi (\partial M) 
= 
0$ then $\delta\chi(\mathcal{H}) = -\delta \sum_{i =1 }^{N} \chi (h_{i})$.

\begin{figure}[!htbp]
    \centering
    \includegraphics[width=0.25\textwidth]{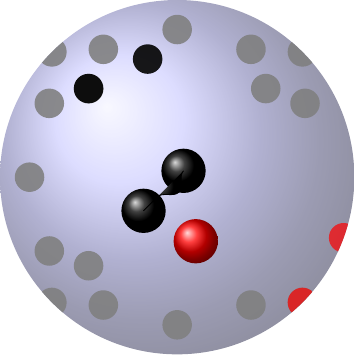}
    \caption{\textit{Each time a BH horizon is formed from gravitational 
collapse (the red 
sphere), two corresponding puncture holes appear on the horizon (red 
disks). Similarly, each time two BH horizons merge (black 
binary), two corresponding puncture holes close up on the apparent horizon 
(black disks). The grey holes indicate an arbitrary number of disk 
punctures excess on the horizon, since only a fraction of the BHs merge.} }
 \label{fig:holographic holes}
\end{figure}

Now, the 
change in the Euler characteristic of BH horizons at every BH formation from 
collapse is $\delta \chi (h) = 2$, while at every merger it is $\delta \chi (h) 
= -2$. Therefore, if   $\delta N_{form}$ BH formations and $\delta 
N_{merg}$ BH mergers 
occur, then the corresponding change of the Euler characteristic of the 
apparent 
horizon will be 
\begin{equation}\label{delta chi apparent horizon}
    \delta \chi(\mathcal{H})= -2 \;(\delta N_{form} - \delta N_{merg}).
\end{equation}
By demanding that the total topology of the causally connected boundaries 
remains constant, each time a BH horizon is formed  the Euler characteristic 
of the apparent horizon decreases by $2$, while each time    two BH 
horizons merge into one  
the Euler characteristic of the apparent horizon  increases by $2$. For 
two-dimensional surfaces the Euler characteristic is given by 
\cite{Farb:2012mcg}
\begin{equation}\label{chi 2D surface}
    \chi = 2 - 2 g - b,
\end{equation}
where $g$ is the genus, the number of handles the surface possesses (torus-like 
 
holes) and $b$ is the number of boundary components  puncture-like holes. An 
interpretation of \eqref{delta chi apparent horizon} according to \eqref{chi 2D 
surface} could be that each time a BH horizon is formed two compensating 
puncture disks appear (open up) on the causal horizon, while  each time 
two BH horizons merge two disk punctures disappear (close up) on the causal 
horizon, as it has been illustrated in Fig.~\ref{fig:holographic holes}.
We mention here that   the connection between the Universe horizon (the largest 
scale of the theory) and small scales is known to hold according to 
the holographic principle  
\cite{tHooft:1993dmi,Susskind:1994vu,Fischler:1998st,Bousso:2002ju,
Horava:2000tb}, with a famous cosmological 
application being the holographic dark energy 
\cite{Li:2004rb,Wang:2016och, Huang:2004ai,Pavon:2005yx,
Wang:2005jx,
Nojiri:2005pu, 
Li:2009bn,  Setare:2008pc,Setare:2008hm, Saridakis:2017rdo, 
Saridakis:2018unr,Drepanou:2021jiv}. In similar lines,  in the following we 
will 
obtain a dark energy sector of topological  nature.

In summary, a first consequence of the above considerations is that  the second 
law of thermodynamics is satisfied by Hawking area theorem 
\cite{Hawking:1971tu}. A second consequence is that the topology of the causal 
horizon becomes dynamical, thus a new term of topological origin appears when 
one derives the Friedmann equations in the spacetime-thermodynamics framework 
for the Wald-Gauss-Bonnet entropy, as we will see in the next subsection.

\subsection{Modified Friedmann equations form Wald-Gauss-Bonnet entropy}

In subsection \ref{Friedmann equations} we showed how the standard Friedmann 
equations can be obtained from the first law of thermodynamics in the case of 
Bekenstein-Hawking entropy that corresponds to general relativity. In the 
present subsection we will 
follow the same procedure  for the case of the Wald-Gauss-Bonnet entropy 
discussed in subsection
\ref{Wald Gauss Bonnet entropy}.

Differentiating the expression of the Wald-Gauss-Bonnet entropy 
\eqref{WGB entropy}, and using \eqref{apparent horizon radius}, we find
\begin{equation}
    \frac{dS}{dt} = - \frac{2 \pi \tilde{r}_{A}^4}{4 G} H \left( \Dot{H} - 
\frac{k}{a^2}\right) + \frac{\pi \tilde{\alpha}}{G} \dot{\chi}(\mathcal{H}).
\end{equation}
Thus, substituting  into the first law of thermodynamics $-dE = T dS$ where 
$dE$ is 
given by \eqref{dE 1st law} and $T$ by \eqref{apparent horizon temperature}, we 
obtain
\begin{equation}
-4 \pi G(    \rho_{m} + p_{m}) =   \left(\Dot{H} - 
\frac{k}{a^2}\right)  -4\tilde{\alpha} \frac{1}{H} \left(H^2 + 
\frac{k}{a^2}\right)^2 
\dot{\chi}(\mathcal{H}).
\label{Hdottime}
\end{equation}
Hence, inserting the conservation equation \eqref{conservation equation} and 
integrating, we finally acquire
\begin{equation}\label{Hmodel}
    H^2 =  \frac{8 \pi G}{3} \rho_{m} + \frac{k}{a^2} + \frac{\Lambda}{3} + 
4\tilde{\alpha} \int_{0}^t \left(H^2 + \frac{k}{a^2}\right)^{2} \dot{\chi}(H) dt.
\end{equation}
Having in mind the discussion of the previous subsection,
if we define the active BH number as the difference between BH formations and 
BH 
mergers, namely
\begin{equation}\label{N active}
    N = N_{form} - N_{merg},
\end{equation}
 we can express $\dot{\chi}(\mathcal{H})$ according to \eqref{delta chi 
apparent horizon}. Thus,  inserting into \eqref{Hmodel} we finally obtain
\begin{equation}\label{Hmodel time}
    H^2 =  \frac{8 \pi G}{3} \rho_{m} + \frac{k}{a^2} + \frac{\Lambda}{3} - 
8\tilde{\alpha} \int_{0}^t \left(H^2 + \frac{k}{a^2}\right)^{2} \frac{dN}{dt}dt.
\end{equation}

Interestingly enough, through the application of the gravity-thermodynamics 
conjecture in the case of Wald-Gauss-Bonnet entropy, we have obtained    
modified   Friedmann equations depending on the BH formations and mergers  and 
on
the GB coupling constant. Note that in the case where the Gauss-Bonnet term is 
absent one recovers the standard cosmological paradigm, which was expected since 
in 
this case   the Wald-Gauss-Bonnet entropy \eqref{WGB entropy} recovers the 
standard Bekenstein-Hawking one,  and thus the Euler characteristic change does 
not have any effect on the entropy and hence on the gravity-thermodynamics 
conjecture. Finally, note that  while Einstein-Gauss-Bonnet theory in four 
dimensions leads to the same field equations with general relativity, 
since the Gauss-Bonnet term in four 
dimensions is a topological invariant,
its 
implementation within the gravity-thermodynamics approach does lead to extra 
terms due to the entropy changes in the horizon caused by the topology changes 
brought about by
the evolution of the Einstein-Gauss-Bonnet black holes.

We can rewrite equations (\ref{Hmodel time}) and (\ref{Hdottime}) as  
\begin{eqnarray}\
\label{Fr1}
&&    H^2 = \frac{8\pi G}{3}(\rho_m+\rho_{DE})\\
&&   H^2+\dot{H} = -\frac{4\pi G}{3}(\rho_m+3p_m +\rho_{DE}+ 3p_{DE})
\label{Fr2}
\end{eqnarray}
by introducing an effective dark energy sector of topological origin, with 
energy density and pressure    defined as
\begin{equation}
    \rho_{DE} = \frac{3}{8 \pi G}\left( \frac{\Lambda}{3} - 8 \tilde{\alpha} 
\int_{0}^t \left(H^2 + \frac{k}{a^2}\right)^{2} \frac{dN}{dt} dt  \right),
\label{rhode}
\end{equation}
and
\begin{eqnarray}
  &&
  \!\!\!\!\!\!\!\!\!\!\!\!\!\!\!\!\!\!
  p_{DE} = -\frac{1}{4\pi G}\left[ \frac{\Lambda}{2}+ 8\tilde{\alpha} 
\frac{1}{H} \left(H^2 + 
\frac{k}{a^2}\right)^2\frac{dN}{dt}\right.\nonumber\\
&&\left. \ \ \ \ \ \ \ \ \  \ \
- 
12 \tilde{\alpha} 
\int_{0}^t \left(H^2 + \frac{k}{a^2}\right)^{2} \frac{dN}{dt} dt   \right].
\end{eqnarray}
Additionally, we   define the effective 
dark-energy equation-of-state parameter as
$w_{DE}\equiv \frac{p_{DE}}{\rho_{DE}}$, leading to
\begin{equation}\label{w time}
    w_{DE} = -1 + \frac{2 \tilde{\alpha} \left( H^2 + \frac{k}{a^2}\right)^2 
\frac{dN}{dt}}{H 
\left( \frac{\Lambda}{4} - 6 \tilde{\alpha} \int_{0}^{t} \left( H^2 + 
\frac{k}{a^2}\right)^2 
\frac{dN}{dt}dt \right)}.
\end{equation}
Lastly, from (\ref{Fr1}),(\ref{Fr2}), and assuming that the matter sector is 
conserved, i.e.
$
\dot{\rho}_{m}+3H(\rho_{m}+p_{m})=0$,
we acquire that the effective dark energy sector is conserved too, namely 
$
\dot{\rho}_{DE}+3H(\rho_{DE}+p_{DE})=0$.

We close this section by mentioning that   in the  standard 
gravity-thermodynamics approach presented in equations 
(\ref{apparent horizon radius})-(\ref{1_FriedmannLCDM}) above, 
 the 
cosmological constant arises as an 
integration constant. In  the case of the modified Friedmann equations form 
Wald-Gauss-Bonnet entropy we also have this term, namely in the 
effective dark energy (\ref{rhode}) we do have the cosmological constant 
(arising as
integration constant). Thus, the Wald-Gauss-Bonnet extra terms 
  are just corrections on top of the cosmological constant $\Lambda$,   
corrections that aim to improve the behavior comparing 
to $\Lambda$CDM scenario.  
Lastly, note that  in our model the dark matter sector enters in the 
same way as it does in standard cosmology, namely it is a separate sector that 
it is added in the model by hand (actually by applying the 
gravity-thermodynamics conjecture one starts in (\ref{dE 1st law}) by 
considering that dark 
and baryonic matter exist in the  Universe), i.e  dark matter is not 
related to the black holes. Hence, dark matter perturbations and clustering 
will be the same as in standard cosmology.\\

\section{\label{Cosmic evolution}Cosmic evolution}

In the previous section we applied the gravity-thermodynamics conjecture in the 
case of Wald-Gauss-Bonnet entropy, and we obtained modified Friedmann equations 
and an effective dark energy sector of topological origin, dependent on the 
black hole formation and merger. Therefore, we can now proceed to the 
investigation 
of the cosmological implications.

For convenience we use the redshift $z$ as the independent variable, 
defined as $1 + z = a_0/a$, and we set the current scale factor to $a_0 = 1$. 
Additionally, we will focus on dust matter (namely with $p_m=0$), which implies
 $\rho_m=\rho_{m0} (1 + z)^{3}$, with $\rho_{m0}$   the matter energy 
density at present. As usual, we introduce the dimensionless density parameters
\begin{equation}\label{Omega_DE}
    \Omega_{DE} = \frac{8 \pi G}{3 H^{2}} \rho_{DE},
\end{equation}
and   
\begin{equation}\label{Omega m}
    \Omega_m = \frac{8 \pi G}{3 H^2} \rho_m.
\end{equation}
Inserting the above into (\ref{rhode}) 
 we obtain
\begin{equation}\label{rho z}
    \rho_{DE}(z)  = \frac{3}{8 \pi G}   
    \left\{ \frac{\Lambda}{3}  - 8 \tilde{\alpha} \int_{z_{i}}^z [H^2 + k 
(1+z)^2]^{2} \frac{dN}{dz} dz  \right\},
\end{equation}
while (\ref{w time}) gives 
\begin{equation}\label{w z}
     w_{DE}\left(z\right)= -1   -\frac{2 \tilde{\alpha} \left[ H^2 + k 
(1+z)^2\right]^2 (1+z) 
\frac{dN}{dz}}{\frac{\Lambda}{4} - 6 \tilde{\alpha} \int_{z_{i}}^z \left[ H^2 + k 
(1+z)^2\right]^2 \frac{dN}{dz} dz}.
\end{equation}
Lastly, we can introduce the deceleration parameter  $q \equiv -1 - 
\dot{H}/H^2$, which  in terms of redshift becomes
\begin{equation}\label{q z}
    q(z) = -1 + \frac{(1+z)}{H(z)} \frac{d{H(z)}}{dz}.
\end{equation}

As we see, the effective dark energy density depends on the black hole 
formation and merger rates, namely on $ \frac{dN}{dz}$. Thus, in the following 
subsection we provide an estimation for its value. 

\subsection{Black hole formation and merger rates}

Let us estimate the rate of the number of BHs that form and merge per redshift. 
In order to achieve that we need to estimate independently the  black-hole 
formation rate and the  black-hole  merger rate.

\subsubsection{Estimating the black-hole formation rate  from the star 
formation 
rate }

\begin{figure}[!htbp]
    \centering
    \includegraphics[width=0.44\textwidth]{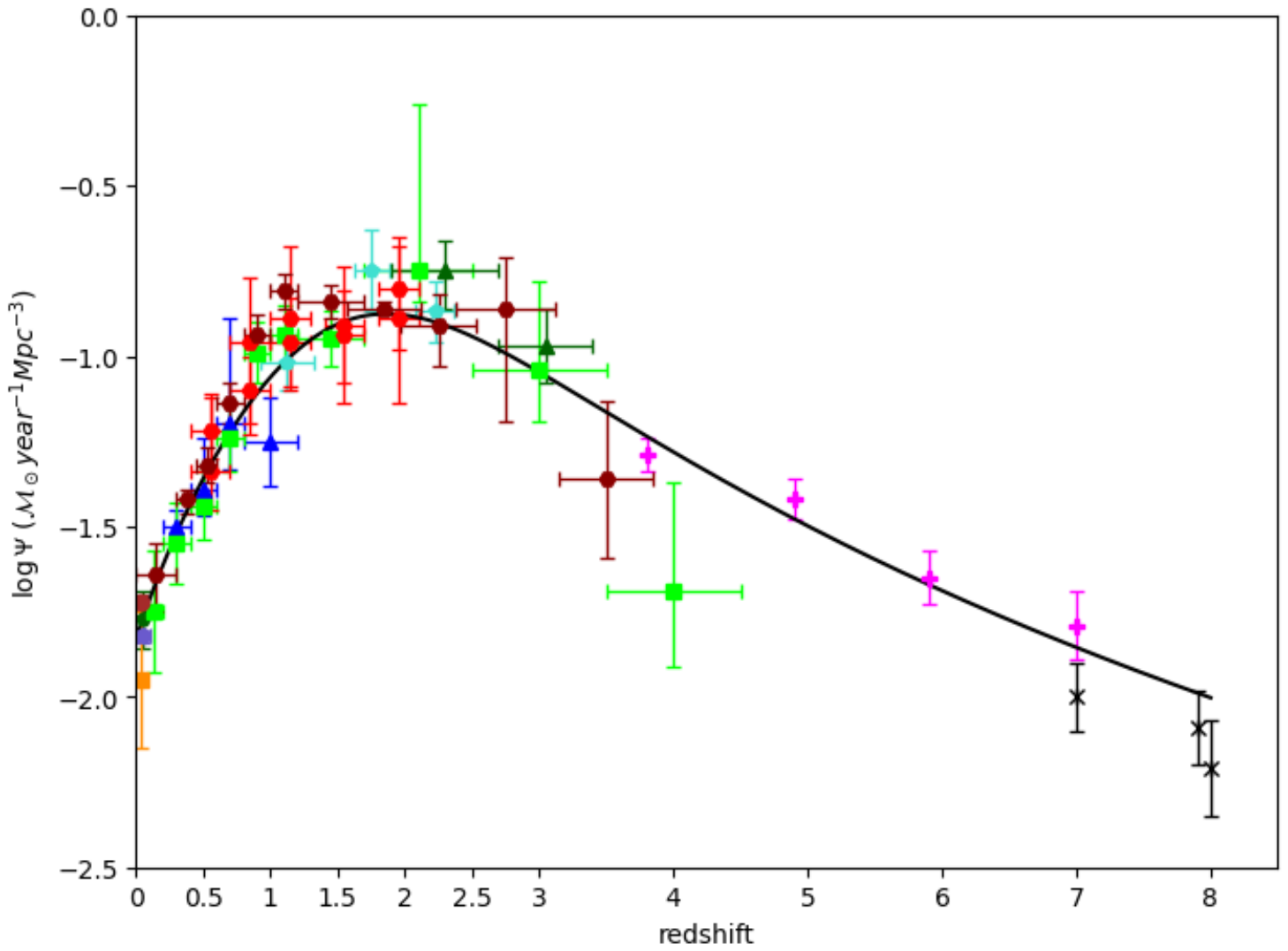}
    \caption{\textit{The history of cosmic star formation rate from 
\cite{Madau:2014bja}. On top of the 
data points, the black solid curve represents the best fit to the data, namely 
the  Madau-Dickinson   form (\ref{SFRfir}).}}
    \label{fig:SFH}
\end{figure}

Since  black holes typically evolve from massive stars, it is   commonly 
assumed that the formation rate of BHs (BHFR) is proportional to the cosmic 
star formation rate (SFR) 
\cite{Iacovelli:2022bbs,Gupta:2023lga,Lehoucq:2023zlt}. The cosmic star 
formation rate density 
is presented in Fig. \ref{fig:SFH}, and its   best fit form is given by 
Madau and 
Dickinson \cite{Madau:2014bja} as
\begin{equation}
   \psi(z) = 0.015 \frac{\left(1 + z\right)^{2.7}}{1 + \left[\left(1 + 
z\right)/2.9\right]^{5.6}}\;M_\odot \text{year}^{-1} Mpc^{-3}.
\label{SFRfir}
\end{equation}

To estimate the BHFR from the SFR, we will consider the fraction of stellar 
mass 
that ends up as BH progenitors. This fraction depends on the initial mass 
function (IMF) and the mass range of stars that collapse into BHs. Assuming a 
Salpeter IMF \cite{Salpeter:1955it}, which gives \( \xi(m) \propto m^{-2.35} 
\), and considering stars with initial masses \( m > 25 \, M_\odot \) as BH 
progenitors, the fraction \( f_{\text{BH}} \) of stellar mass forming BH 
progenitors is
\begin{equation}\label{fBH IMF}
f_{\text{BH}} = \frac{\int_{25\,M_\odot}^{\infty} m \, \xi(m) \, 
dm}{\int_{0.1\,M_\odot}^{\infty} m \, \xi(m) \, dm},
\end{equation}
and it represents the fraction of total stellar mass that goes into stars 
massive enough to eventually form BHs. According to various studies, \( 
f_{\text{BH}} \) ranges from approximately 0.001 to 0.05 
\cite{Heger:2002by,Fryer:2011cx}.

The average mass of progenitor stars that will evolve to form BHs, namely \( 
\langle 
m_{\text{prog}} \rangle \), is typically in the range of \( 25 \, M_\odot \) to 
\( 40 \, M_\odot \) \cite{Woosley:2002zz,Sukhbold:2015wba}. Thus, the number 
rate 
density of BHs formed per unit volume per unit time, \( 
\dot{\rho}_{\text{BH}}(z) \), is given by
\begin{equation}
   \dot{\rho}_{\text{BH}}(z) = f_{\text{BH}} \frac{ \psi(z)}{\langle 
m_{\text{prog}} \rangle}.
\end{equation}
Hence, the number rate of BHs that form inside the apparent horizon of 
volume 
$\tilde{V}_{A}=\frac{4 \pi}{3} \tilde{r}_{A}^3$, will be $\dot{N}_{\text{BH}} = 
\dot{\rho}_{\text{BH}}   \tilde{V}_{A} $, and thus  for a flat Universe 
from 
\eqref{apparent horizon radius} we  have $\tilde{r}_{A} = 1 / H$, which leads to
\begin{equation}\label{dotN BH}
    \dot{N}_{\text{BH}}(z) = \frac{4 \pi}{3}  f_{\text{BH}} \frac{ 
\psi(z)}{\langle m_{\text{prog}} \rangle H^{3}(z)}.  
\end{equation}
Finally,  expressing the number rate of BHs formation per redshift 
instead of   time,  using
  $|dt/dz| = 1 / H(z)( 1 + z )$, yields
\begin{equation}\label{dNBH/dz}
    \frac{dN_{\text{BH}}(z)}{dz} = \frac{4 \pi}{3}  f_{\text{BH}} \frac{ 
\psi(z)}{\langle m_{\text{prog}} \rangle (1 + z) H^{4}(z)}.
\end{equation}

\subsubsection{Estimating the  black-hole  merger rate from the black-hole 
formation rate}

The binary black hole  merger rate (BHMR)  can also be assumed to be 
proportional 
to SFR. Nevertheless, ambiguities arise in this simplified approach as the 
formation efficiency of compact object is metallicity-dependent and BH 
formation and binary black hole merging may occur with a significant delay 
relative to the star formation epoch \cite{Boesky:2024wks}. 
We are going to estimate approximately the BHMR
from the BHFR by considering as main factors the fraction of massive stars that 
are in binary systems and the   merger efficiency.  Observations suggest that 
a significant fraction of massive stars are in binaries, which is estimated 
around 
\( f_{\text{bin}} \approx 0.7 \) \cite{Sana:2012px}. Nevertheless, not all 
binary black holes will merge within the age of the Universe. The merger 
efficiency \( f_{\text{merge}}\) accounts for the fraction of binary black 
holes 
that   will eventually merge  within a Hubble time. This efficiency depends 
on factors such as the initial separation and eccentricity of the binary 
system, 
and it is estimated to be between 0.01 and 0.1 
\cite{Belczynski:2016obo,Eldridge:2016ymr,Dominik:2012kk}. Therefore, the black 
hole merger rate with respect to redshift  will be
\begin{equation}
    \frac{dN_{\text{BHMR}}(z)}{dz} = f_{\text{bin}} \times f_{\text{merge}} 
\times 
\frac{dN_{\text{BH}}(z)}{dz}, 
\end{equation}
which using \eqref{dNBH/dz} yields
\begin{align}\label{dNBHmerge/dz}
    \frac{dN_{\text{BHMR}}(z)}{dz} =& \frac{4 \pi}{3}  f_{\text{BH}} \times 
f_{\text{bin}} \times f_{\text{merge}}\times \\ \nonumber
    &\frac{ \psi(z)}{\langle m_{\text{prog}} \rangle H^{4}(z) (1 + z) }.
\end{align}

\subsection{Specific examples}

Having estimated the black-hole formation and merger rates in  \eqref{dNBH/dz} 
and \eqref{dNBHmerge/dz}, we can insert  them in the derivative of  \eqref{N active} pre redshift and obtain the 
 expression for the rate of the active number of BHs per redshift 
inside the apparent horizon as
\begin{align}\label{active dN/dz}
    \frac{dN(z)}{dz} =& C  \frac{ \psi(z)}{  H^{4}(z) (1 + z)},
\end{align}
where we have defined  the constant 
\begin{equation}
    C \equiv \frac{4 \pi}{3} \frac{ \left( 1 - f_{\text{bin}}  \times 
f_{\text{merge}}\right)  f_{\text{BH}}}{\langle m_{\text{prog}} \rangle}.
\end{equation}
Hence, inserting (\ref{active dN/dz}) into  
\eqref{Fr1},\eqref{rho z} and \eqref{w z} we acquire
\begin{equation}\label{SFR Hmodel z}
    H^2 = H_0^2 \Omega_{m0} (1 + z)^{3}  + \frac{\Lambda}{3} - 8 \tilde{\alpha} C 
\int_{z_{i}}^z  \frac{ \psi(z)}{ (1 + z)} dz,
\end{equation}
\begin{equation}\label{SFR rho z}
    \rho_{DE}(z) = \frac{3}{8 \pi G}\left( \frac{\Lambda}{3} - 8 \tilde{\alpha} C 
\int_{z_{i}}^z  \frac{ \psi(z)}{ (1 + z)} dz\right),
\end{equation}
and
\begin{equation}\label{SFR w z}
    w_{DE}\left(z\right)=-1-\frac{2 \tilde{\alpha} C \psi(z) }{\frac{\Lambda}{4} - 6 
\tilde{\alpha} C \int_{z_{i}}^z  \frac{ \psi(z)}{ (1 + z)} dz},
\end{equation}
with $H_0$ the current value of the Hubble function. 
As we see, apart from the constants, the only dynamical function that enters 
the equations is the star formation rate $ \psi(z)$, which is given by 
(\ref{SFRfir}). The integral  can be evaluated   
in terms of the hypergeometric function ${}_2F_1(a, b; c; z)$, and gives
\begin{align}
    \int & \frac{ \psi(z)}{ (1 + z)} dz = 0.37037 \cdot (1 + z)^{2.7} \\ 
\nonumber
    & {}_2F_1\left(0.482143,\, 1.0;\, 1.48214;\, -0.00257378 \cdot (1 + 
z)^{5.6}\right). 
\end{align}
Finally, concerning the involved parameters, in Table~\ref{tab:parameter table} 
we display the possible values according to the literature.

\begin{table}[!htbp]
\caption{\label{tab:parameter table}%
Range of the involved parameters according to the literature.}
\begin{ruledtabular}
\begin{tabular}{lcr}
Parameter & Value & Reference \\
\hline
\( f_{\text{BH}} \) & 0.1\% to 5\% & \cite{Heger:2002by,Fryer:2011cx} \\
\( \langle m_{\text{prog}} \rangle \) & 25 to 40 \( M_\odot \) & 
\cite{Woosley:2002zz,Sukhbold:2015wba} \\
\( f_{\text{merge}} \) & 1\% to 10\% & 
\cite{Belczynski:2016obo,Eldridge:2016ymr,Dominik:2012kk} \\
\( f_{\text{bin}} \) & 50\% to 80\% & 
\cite{Sana:2012px,Moe:2016tmr,Duchene:2013uma} \\
\end{tabular}
\end{ruledtabular}
\end{table}

\subsubsection{\texorpdfstring{$\tilde{\alpha} > 0$}{tilde alpha > 0} case}

We   start our analysis by examining the case where the GB coupling constant 
$\tilde{\alpha}$ is positive.
\begin{figure}[!]
             \includegraphics[width=0.34\textwidth]{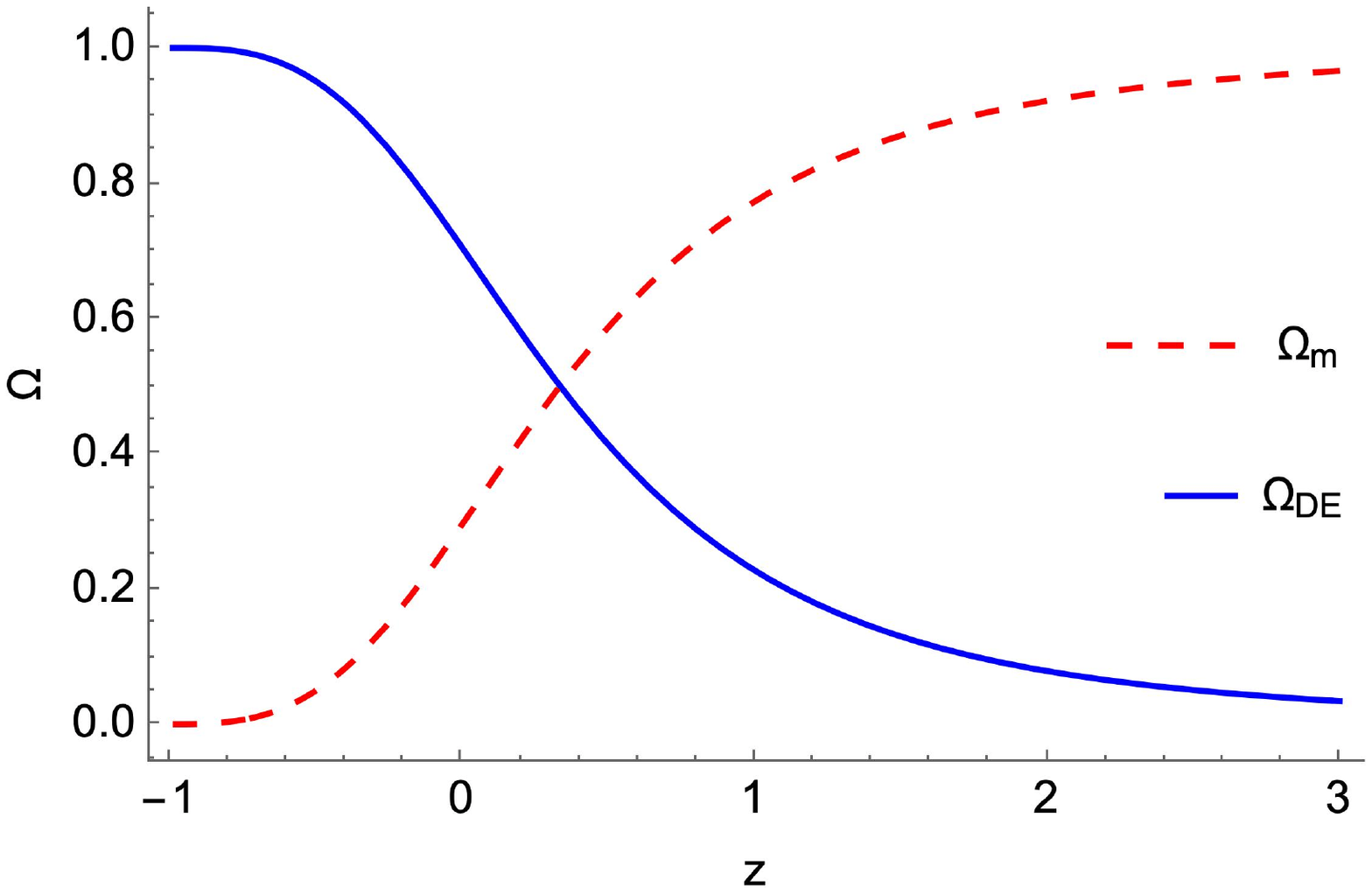}
        \includegraphics[width=0.37\textwidth]{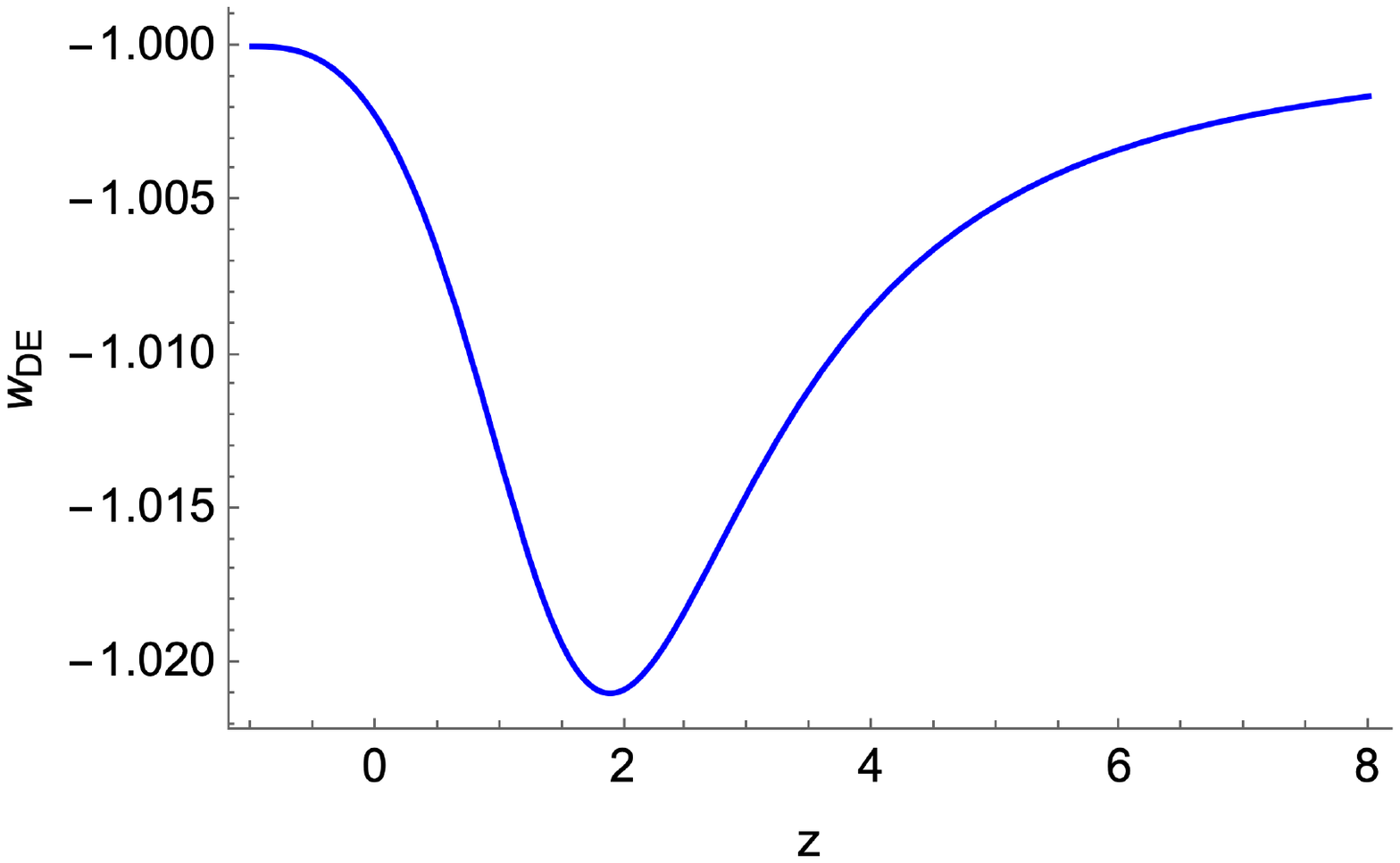}
           \includegraphics[width=0.34\textwidth]{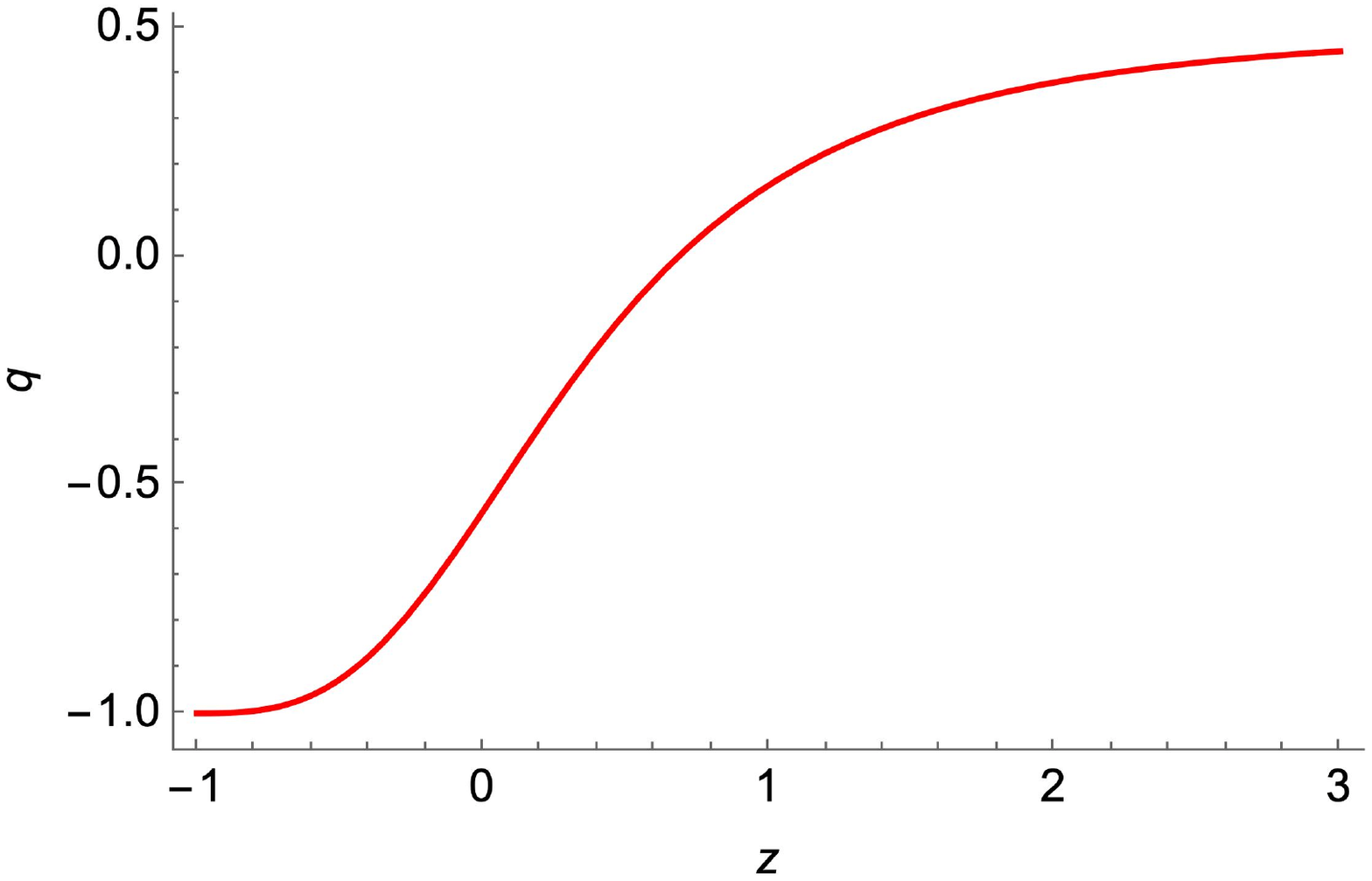}
    \caption{\textit{Upper graph: The evolution of the dimensionless dark 
energy 
parameter $\Omega_{DE}$ (blue-solid) and the corresponding matter density 
parameter $\Omega_m$ (red-dashed) as a function of redhsift, for the modified 
scenario with Wald-Gauss-Bonnet entropy, with $\tilde{\alpha}>0$. Middle graph: 
The evolution of the 
dark energy equation of state parameter $w_{DE}$. Lower graph: The evolution 
of the 
deceleration parameter $q$. In all graphs we have used the models parameters 
$\tilde{\alpha}= 10^5$ (in $H_0$ units), $f_{BH}=0.025$, $m_{prog}=30 
M_{\odot}$, 
$f_{bin}=0.65$, $f_{merge}=0.05$ and we have implemented $\Omega_{DE0}=0.69$.}}
    \label{fig:a>0_three_plots}
\end{figure}
We elaborate  the cosmological equations numerically, imposing   
$\Omega_{DE}(z=0)\equiv\Omega_{DE0}\approx0.69$ and   
$\Omega_m(z=0)\equiv\Omega_{m0}\approx0.31$  as required by 
observations
\cite{Planck:2018vyg}. 
In the upper graph of Fig. \ref{fig:a>0_three_plots} we present the evolution 
of 
the dimensionless density parameters $\Omega_{DE}(z)$ and $\Omega_m(z)$. We 
observe that we can acquire the usual thermal history of the Universe, the 
standard sequence of matter and dark energy epochs, while in the asymptotic 
future ($z \rightarrow -1 $) the Universe results to be completely dominated by 
dark energy. In the middle graph we draw the evolution of the 
dark-energy equation-of-state parameter  $w_{DE}(z)$, where we can see that it 
lies slightly in the phantom regime for intermediate 
redshifts, well inside the observational bounds \cite{Planck:2018vyg}, while it 
approaches $w_{DE}\rightarrow -1$ in the distant past and in the asymptotic 
future. Finally, in the last graph of Fig. \ref{fig:a>0_three_plots} we depict 
the deceleration parameter $q(z)$, where we can see 
that the transition from deceleration to acceleration occurs at $z_{tr} \approx 
0.6$ in agreement with observations.

\begin{figure}[!htbp]
    \centering
    \includegraphics[width=0.40\textwidth]{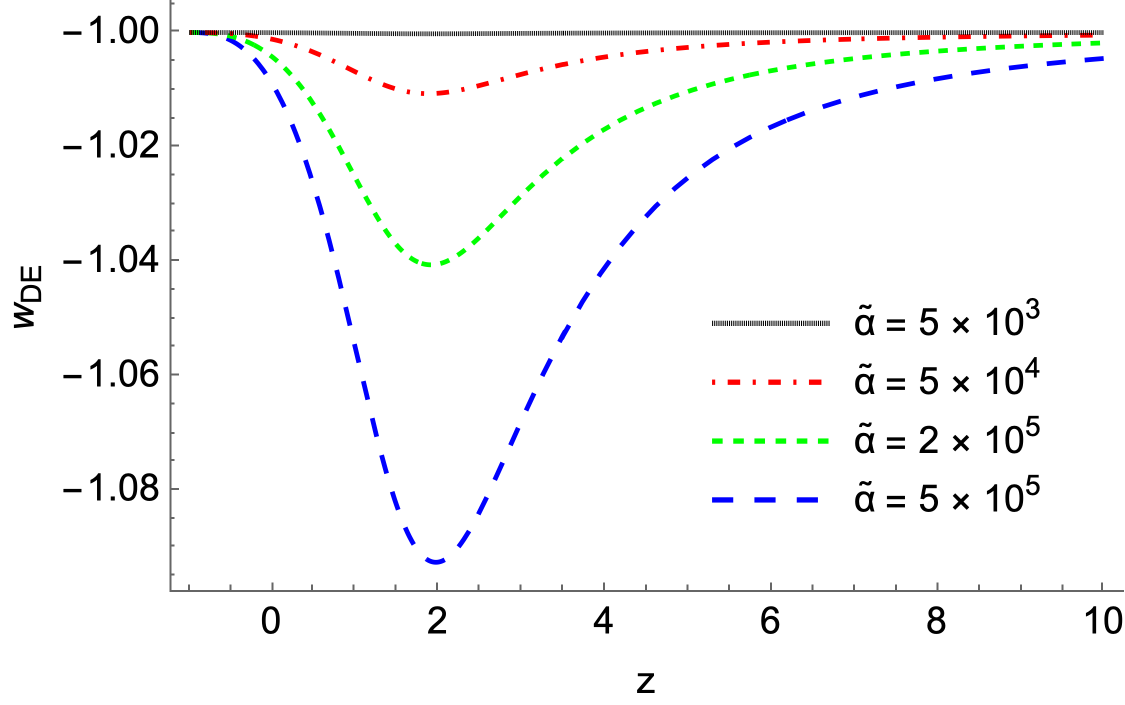}
    \caption{\textit{The evolution of the effective dark-energy 
equation-of-state parameter $w_{DE}$ for various values of positive GB coupling 
constant 
$\tilde{\alpha}$ in $H_0$ units. The other model parameters used in the 
calculation are 
$f_{BH}=0.025$, $m_{prog}=30 M_{\odot}$, $f_{bin}=0.65$, $f_{merge}=0.05$,
and 
we have imposed $\Omega_{DE0}=0.69$. In   all cases the density parameters 
  exhibit similar behaviors with those presented in the upper graph of 
Fig.~\ref{fig:a>0_three_plots}.}} 
    \label{fig:wDEplus alpha}
\end{figure}

We proceed by studying the effect of the positive GB coupling constant 
$\tilde{\alpha}$ on the   dark-energy equation-of-state parameter. In 
Fig.~\ref{fig:wDEplus alpha} we can see that for small values, namely for 
$0\leq\tilde{\alpha} 
\leq 10^3$  in $H_0$ units, the scenario coincides with $\Lambda$CDM paradigm. 
Nevertheless, as the 
value of $\tilde{\alpha}$ becomes larger, $w_{DE}(z)$ enters deeper into the 
phantom regime peaking at $z\approx 2$, while its behavior in the distant past 
and asymptotic future still resembles $\Lambda$CDM one.

\begin{figure}[!htbp]
    \centering
    \includegraphics[width=0.40\textwidth]{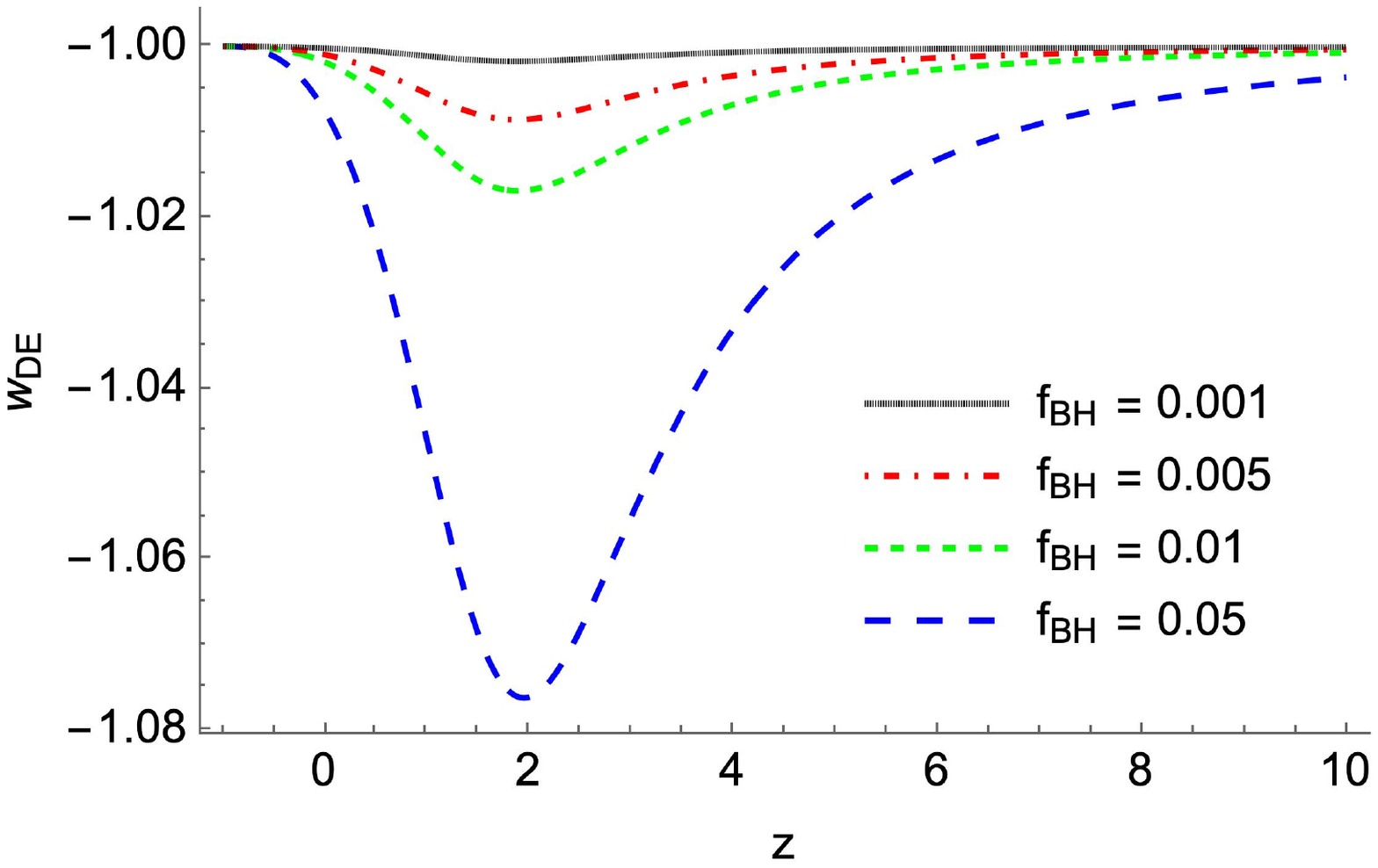}
    \caption{\textit{The evolution of the effective dark-energy 
equation-of-state parameter $w_{DE}$ for various values of the fraction of 
stars 
that form 
black holes $f_\text{BH}$. The other model parameters used in the calculation 
are $\tilde{\alpha}= 2 \times 10^5$  in $H_0$ units, $m_{prog}=30 M_{\odot}$ , 
$f_{bin}=0.65$, 
$f_{merge}=0.05$  and we have imposed $\Omega_{DE0}=0.69$. In all   cases the 
density parameters 
  exhibit similar behaviors with those presented in the upper graph of 
Fig.~\ref{fig:a>0_three_plots}.}}
    \label{fig:wDEplus fBH}
\end{figure}

In order to examine the effect of the   BH formation factor $f_{BH}$ on 
$w_{DE}(z)$, in Fig. \ref{fig:wDEplus fBH} we demonstrate its evolution for  
various values of $f_{BH}$. We can see that for the smallest value 
according to the literature, i.e. $f_{BH}=0.001$, the model tends to 
$\Lambda$CDM 
behavior. As $f_{BH}$ increases $w_{DE}(z)$ enters deeper 
into the phantom regime and remains inside observational bounds 
\cite{Planck:2018vyg} for the highest estimated value $f_{BH}=0.001$. 
Finally,  in the distant past and asymptotic future  we obtain 
$w_{DE}\rightarrow -1$, 
independently of the parameter value. 
\begin{figure}[!htbp]
    \centering
    \includegraphics[width=0.40\textwidth]{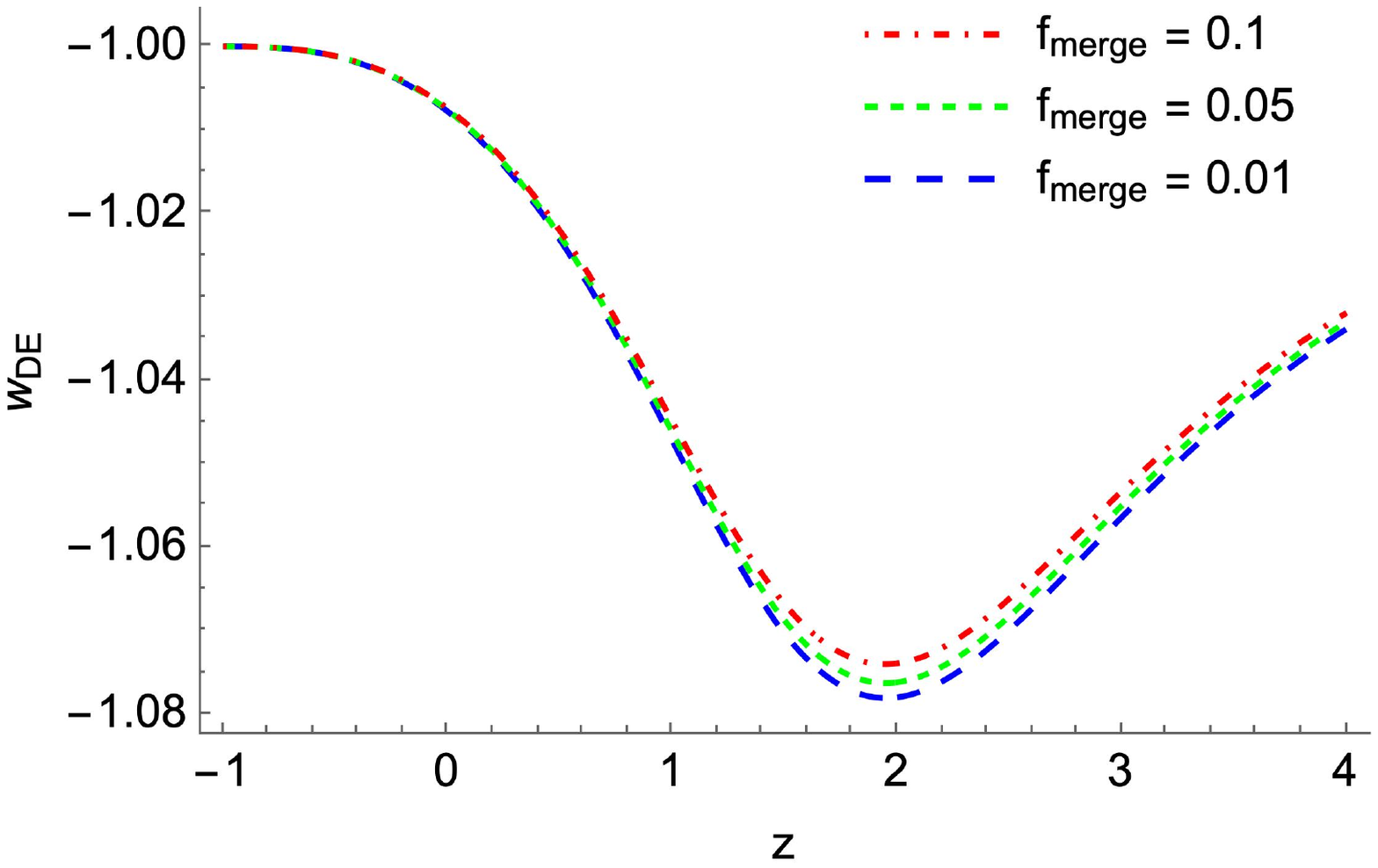}
    \caption{\textit{The evolution of the effective dark-energy 
equation-of-state parameter $w_{DE}$ for various values of the 
fraction of BHs that 
eventually merge $f_\text{merge}$. The other model parameters used in the 
calculation are
    $\tilde{\alpha}= 2 \times 10^5$  in $H_0$ units, $f_{BH}=0.025$, 
$m_{prog}=30 M_{\odot}$, 
$f_{bin}=0.65$, and we have imposed $\Omega_{DE0}=0.69$. In all   cases the 
density parameters 
  exhibit similar behaviors with those presented in the upper graph of 
Fig.~\ref{fig:a>0_three_plots}.} }
    \label{fig:wDEplus merge}
\end{figure}

Finally, we examine the behavior of $w_{DE}(z)$ with respect to the estimated 
range of
values of the BH merging parameter $f_{merge}$. In Fig.~\ref{fig:wDEplus merge} 
one can see that    
$w_{DE}(z)$ increases slightly  as $f_{merge}$ increases, however its effect 
is minor comparing to the previous parameters. Again, in the distant past and 
asymptotic future 
we have $w_{DE}\rightarrow -1$, independently of the parameter value.

\subsubsection{\texorpdfstring{$\tilde{\alpha} < 0$}{tilde alpha < 0} case}

We   now present the cosmological   behavior for negative GB coupling 
constant $\tilde{\alpha}<0$.  
\begin{figure}[!htbp]
           \includegraphics[width=0.34\textwidth]{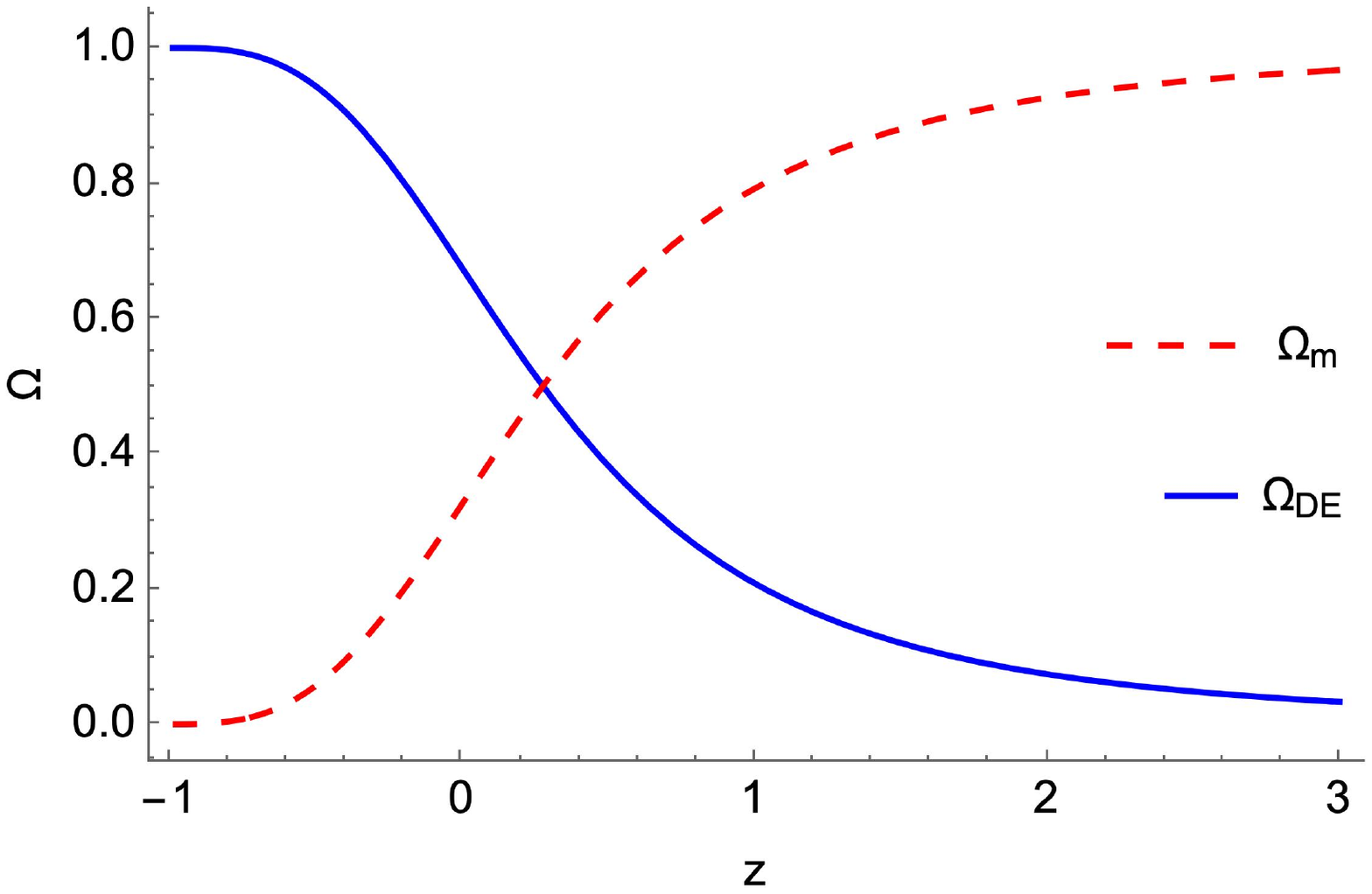}
        \includegraphics[width=0.37\textwidth]{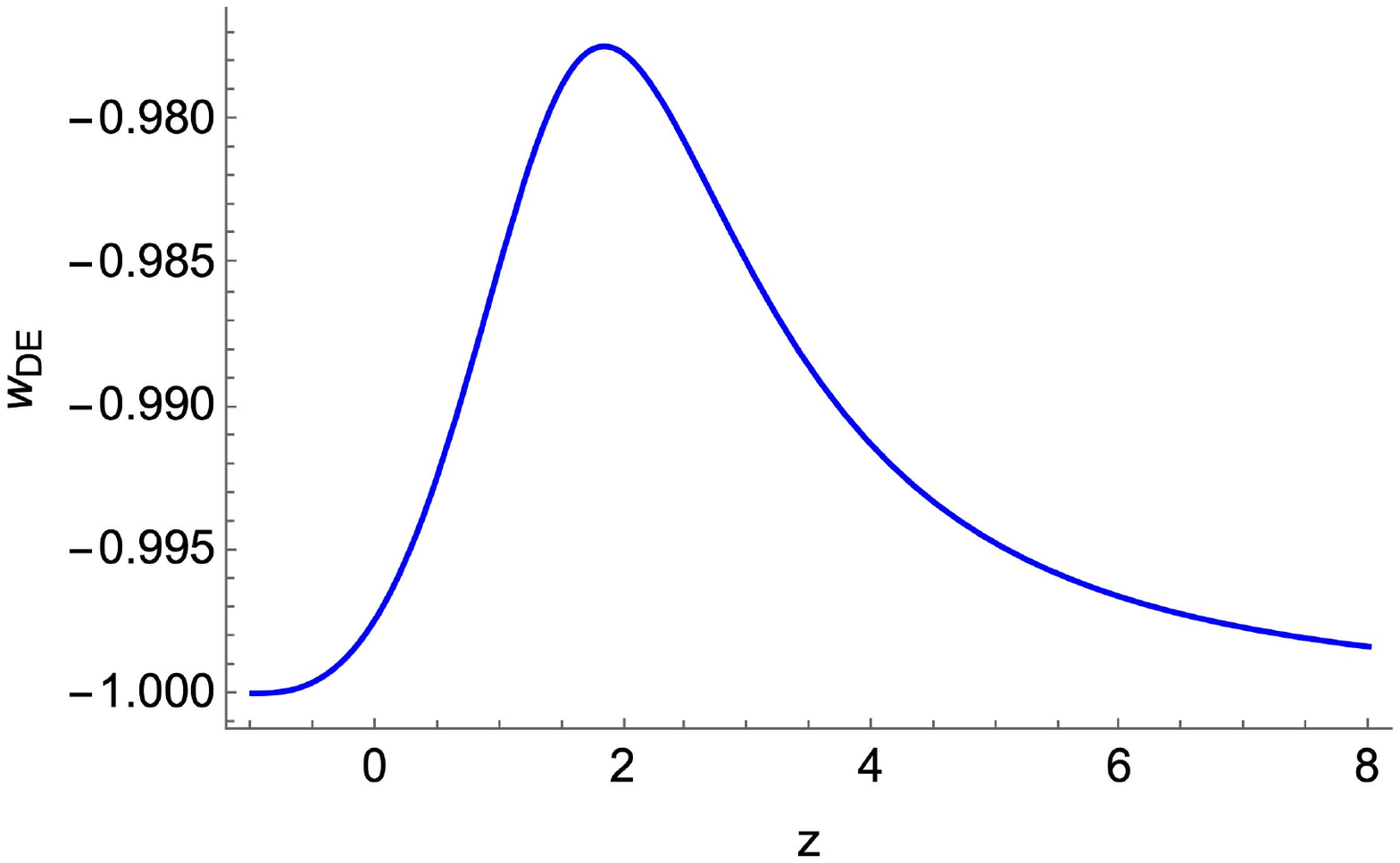}
        \includegraphics[width=0.34\textwidth]{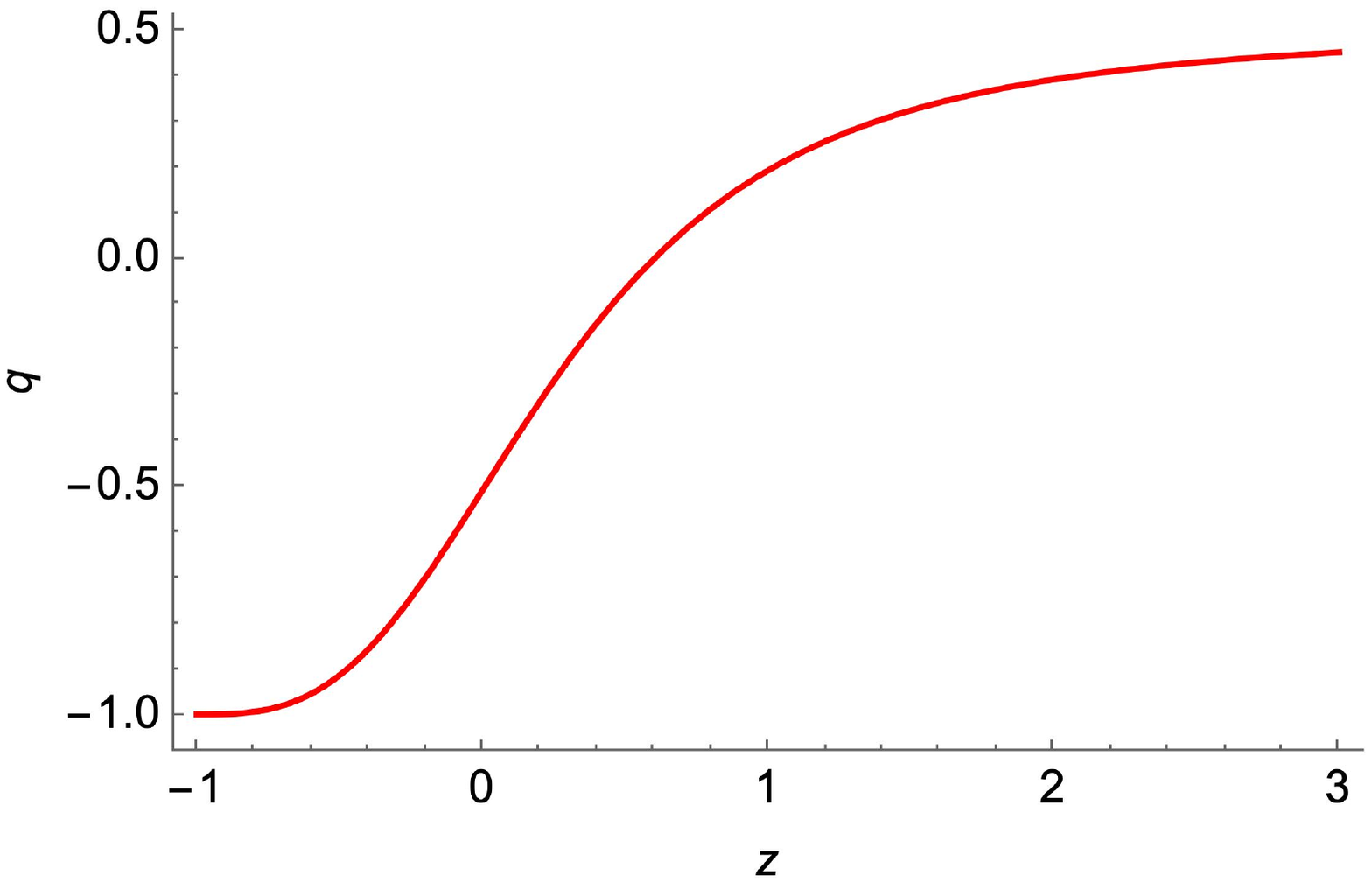}
    \caption{\textit{Upper graph: The evolution of the dimensionless dark 
energy 
parameter $\Omega_{DE}$ (blue-solid) and the corresponding matter density 
parameter $\Omega_m$ (red-dashed) as a function of redhsift, for the modified 
scenario with Wald-Gauss-Bonnet entropy, with $\tilde{\alpha}<0$. Middle graph: 
The evolution of the 
dark energy equation of state parameter $w_{DE}$. Lower graph: The evolution 
of the 
deceleration parameter $q$. In all graphs we have used the models parameters 
$\tilde{\alpha}=-10^5$ (in $H_0$ units), $f_{BH}=0.025$, $m_{prog}=30 
M_{\odot}$, 
$f_{bin}=0.65$, $f_{merge}=0.05$ and we have implemented 
$\Omega_{DE0}=0.69$. } }
    \label{fig:a<0_three_plots}
\end{figure} 
In the upper graph of Fig.~\ref{fig:a<0_three_plots} we present the evolution 
of 
$\Omega_{DE}(z)$ and $\Omega_{m}(z)$, where one can see that the anticipated 
thermal history of the Universe and the sequence of matter and dark energy 
epochs are acquired. Furthermore, in the asymptotic future 
dark energy dominates completely, leading the Universe to a de Sitter phase. In 
the middle 
graph we   observe that the dark-energy equation-of-state parameter lies in the 
quintessence regime ($w_{DE} > -1$) for small redshifts, peaking at $z \approx 
2$, and 
in the asymptotic future it approaches $w_{DE}\rightarrow -1$. Finally, in the 
lower 
graph we   display the evolution of the deceleration parameter and we can 
see that the transition 
from deceleration to acceleration happens at $z_{tr} \approx 0.6$ as 
required. 
\begin{figure}[!htbp]
    \centering
    \includegraphics[width=0.40\textwidth]{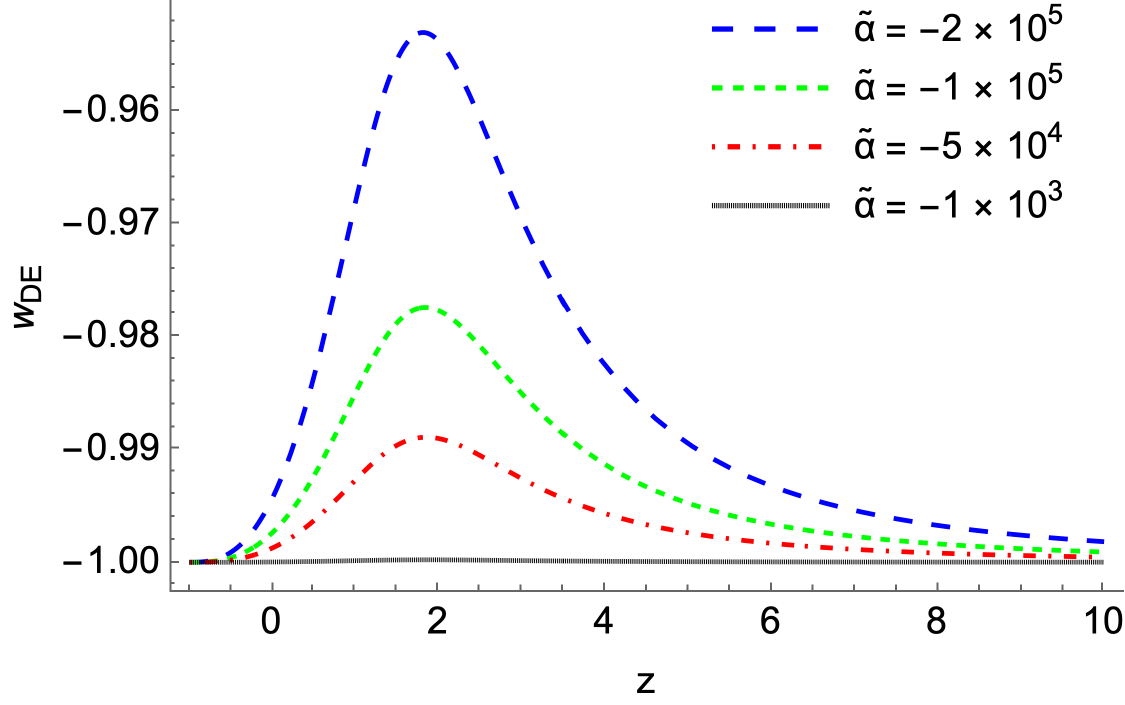}
    \caption{\textit{The evolution of the effective dark-energy 
equation-of-state parameter $w_{DE}$ for various values of negative GB coupling 
constant 
$\tilde{\alpha}$ in $H_0$ units. The other model parameters used in the 
calculation are 
$f_{BH}=0.025$, $m_{prog}=30 M_{\odot}$, $f_{bin}=0.65$, $f_{merge}=0.05$, and 
we have imposed $\Omega_{DE0}=0.69$.  In   all cases the density parameters 
  exhibit similar behaviors with those presented in the upper graph of 
Fig.~\ref{fig:a>0_three_plots}.}}
    \label{fig: wDE minus alpha}
\end{figure}

In Fig.~\ref{fig: wDE minus alpha} we   plot  $w_{DE}$ for different values of 
the negative GB coupling constant. For small absolute values, namely for 
$- 10^3\leq \tilde{\alpha} \leq0$ in $H_0$ units  we retrieve $\Lambda$CDM 
behavior, while when the 
absolute value of the parameter $\tilde{\alpha}$ increases then $w_{DE}$ enters 
upwards to the quintessence regime.
\begin{figure}[!htbp]
    \centering
    \includegraphics[width=0.40\textwidth]{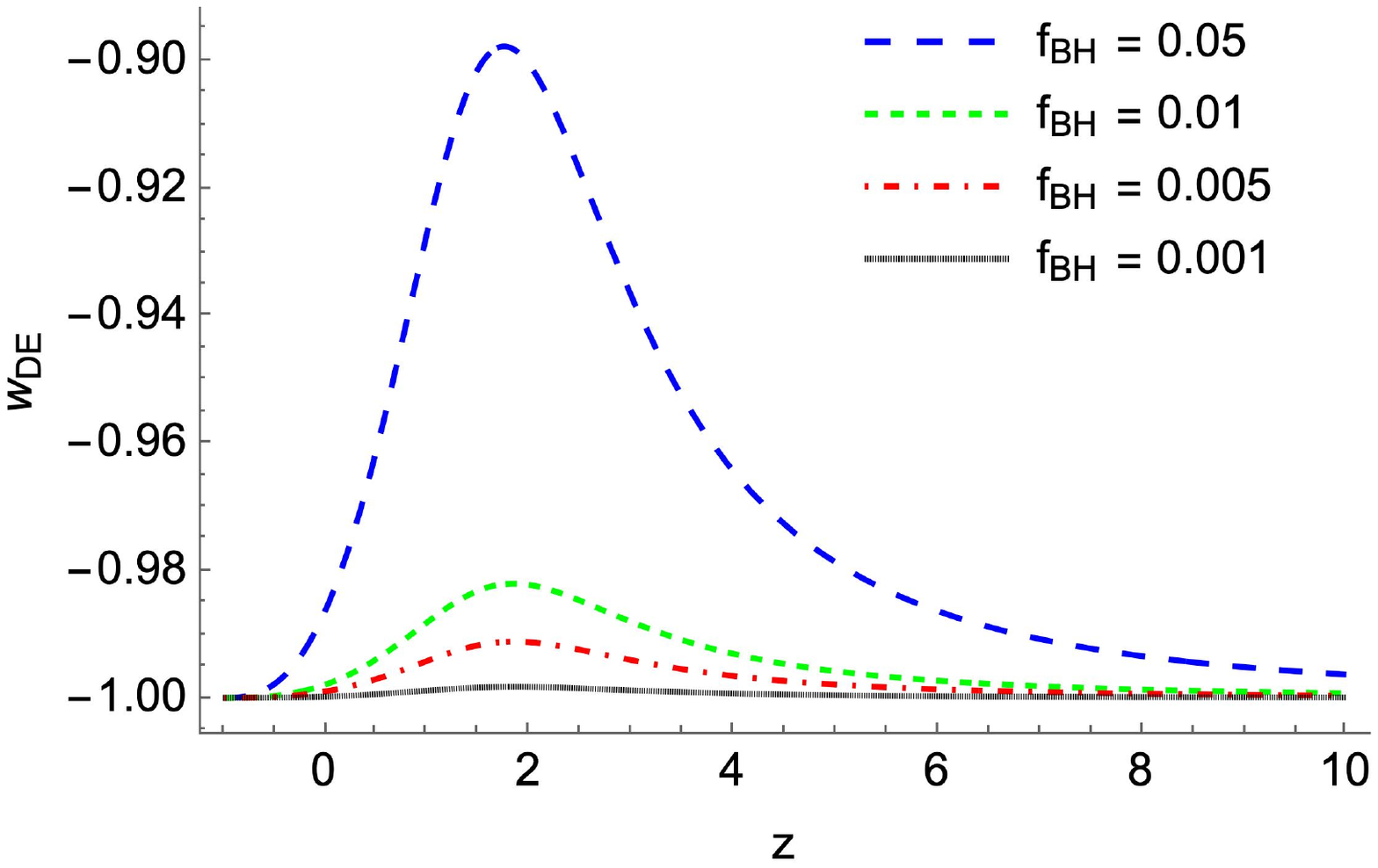}
    \caption{\textit{The evolution of the effective dark-energy 
equation-of-state parameter $w_{DE}$ for various values of the fraction of 
stars 
that form black holes $f_\text{BH}$.  The other model parameters used in the 
calculation are 
$\tilde{\alpha} = - 2 \times 10^5$ in $H_0$ units, $m_{prog}=30 M_{\odot}$, 
$f_{bin}=0.65$, 
$f_{merge}=0.05$, and we have imposed $\Omega_{DE0}=0.69$.  In   all cases the 
density parameters 
  exhibit similar behaviors with those presented in the upper graph of 
Fig.~\ref{fig:a>0_three_plots}.}}
    \label{fig:wDE minus fBH}
\end{figure}
Additionally, in Fig.~ \ref{fig:wDE minus fBH} we draw  $w_{DE}$ for different 
values of   the 
fraction $f_{BH}$ of stars that form BHs. As we can see, for the smallest value 
of the estimated range, i.e. for $f_{BH}= 0.001$, $w_{DE}$ tends to 
$\Lambda$CDM 
behavior, while as the values of $f_{BH}$ increases, $w_{DE}(z)$ enters 
higher to the quintessence regime. However, for the whole allowed $f_{BH}$ 
range showed in Table \ref{tab:parameter table}, $w_{DE}(z)$ remains always 
inside the observational bound \cite{Planck:2018vyg}. 
\begin{figure}[!htbp]
    \centering
    \includegraphics[width=0.40\textwidth]{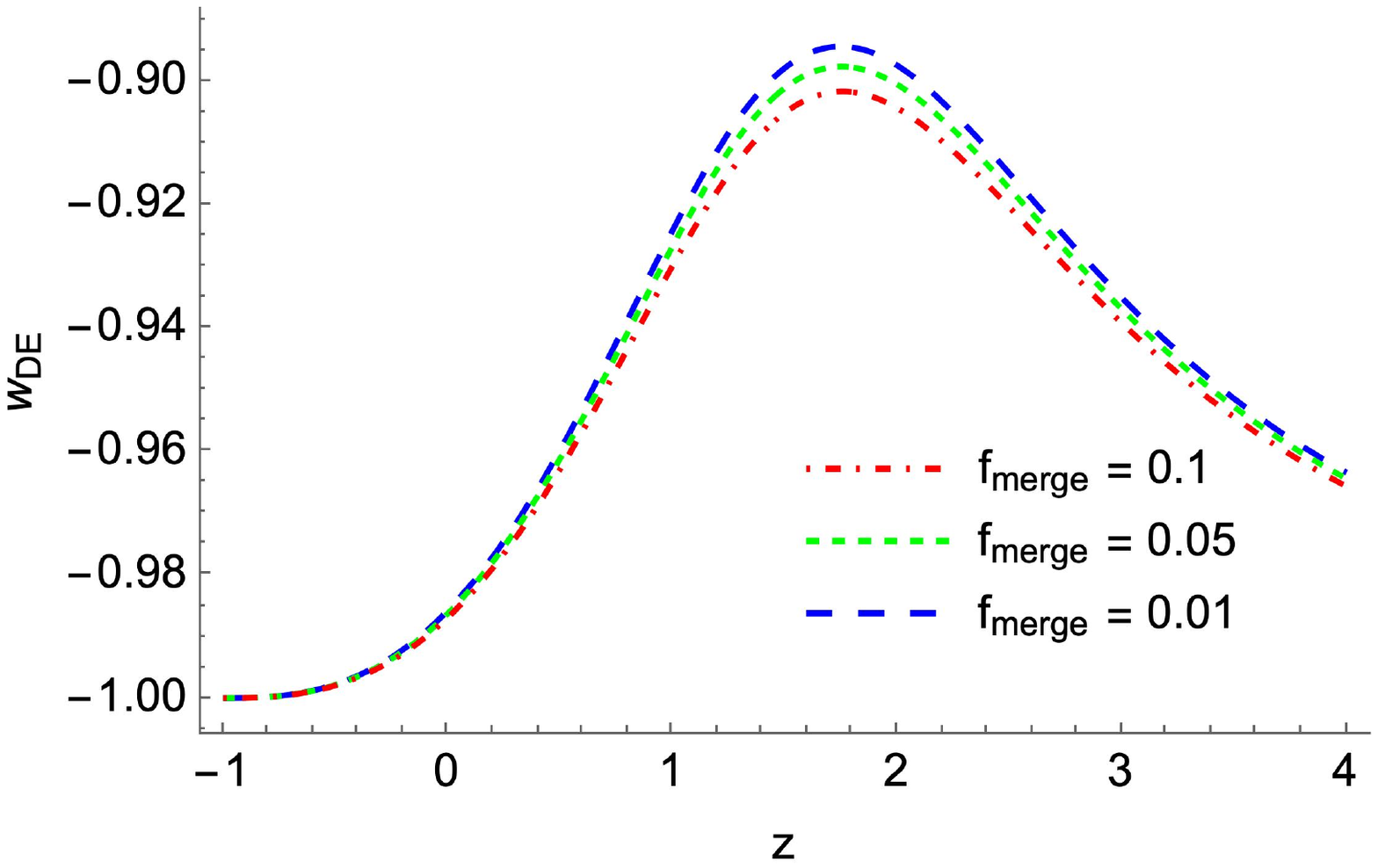}
    \caption{\textit{The evolution of the effective dark-energy 
equation-of-state parameter $w_{DE}$ for various values of the fraction of 
black 
holes that 
eventually merge $f_\text{merge}$. The other model parameters used in the 
calculation are
    $\tilde{\alpha} = - 2 \times 10^5$  in $H_0$ units, $f_{BH}=0.025$, 
$m_{prog}=30 M_{\odot}$, $f_{bin}=0.65$, and we have imposed 
$\Omega_{DE0}=0.69$.  In   all cases the density parameters 
  exhibit similar behaviors with those presented in the upper graph of 
Fig.~\ref{fig:a>0_three_plots}.}}
    \label{fig:wDE minus merge}
\end{figure}
Lastly, in Fig.~\ref{fig:wDE minus merge} we plot $w_{DE}(z)$ for various 
values of the fraction of 
BHs that merge. As we observe,     
$w_{DE}(z)$ decreases slightly  as $f_{merge}$ increases, however its effect 
is minor comparing to the previous parameters.

\section{\label{Conclusions}Conclusions}

In this work we have investigated the Wald-Gauss-Bonnet entropy in the 
framework of spacetime thermodynamics. The latter is a strong conjecture that 
connects gravity and thermodynamics, since by applying the black-hole physics 
in 
the Universe apparent horizon we can obtain the Friedmann equations just 
starting from the first law of thermodynamics. 

We have implemented the above approach in the case of Einstein-Gauss-Bonnet 
gravity and its corresponding Wald-Gauss-Bonnet entropy, which due to the 
Chern-Gauss-Bonnet theorem it is related to the Euler characteristic of the 
Universe topology. Nevertheless, it is known that in the case of the GB 
extension of general relativity, during BH merging the second law is violated. 
In order to remove the  violation we introduced a   topological link between 
the apparent horizon and the   BH horizons, a connection that is known to hold 
according to holographic principle. Hence, through the 
gravity-thermodynamics approach we  extracted  modified Friedmann 
equations, where the new terms depend on the topology changes induced by the 
black-hole formation and merger. Specifically, we obtained an effective, dark 
energy sector of topological origin, which evolves in time according to the 
black-hole formation and 
merger rates.

 In order to investigate the cosmological evolution, we estimated the 
BH formation rate starting from the star formation 
rate, and moreover we estimated  the  black-hole  merger rate from the 
black-hole formation rate. Ultimately, we resulted to a dark-energy energy 
density that depends only on the   cosmic star 
formation rate density per redshift $\psi(z)$, which is parametrized very 
efficiently by the Madau-Dickinson form  (\ref{SFRfir}), with the remaining 
model parameters being   the 
fraction of stars forming 
BHs \( f_{\text{BH}} \), the fraction of black 
holes that 
eventually merge $f_\text{merge}$, the  fraction of massive stars that are in 
binaries \( f_{\text{bin}} \), and the   average mass of progenitor stars that 
will evolve to form BHs \( \langle m_{\text{prog}} \rangle \), as well as    
  the GB coupling constant  $\tilde{\alpha}$. The GB coupling is the only 
completely free parameter, since the other   parameters have 
specific ranges according to the literature. 

We investigated in detail the evolution of the dark energy and matter density 
parameters, of the effective dark-energy equation-of-state parameter, and of 
the deceleration parameter. As we saw,   the Universe evolves according 
to the usual thermal history, with successive dark energy and matter epochs, 
and 
the transition from deceleration to acceleration takes place at $z \approx 0.6$ 
in agreement with observations, while at asymptotically late times the Universe 
results in a  de Sitter phase completely dominated by dark energy. Concerning 
the dark-energy equation-of-state parameter, we showed that it exhibits a 
different behavior according to the sign and value of the  GB coupling 
$\tilde{\alpha}$. For positive values of the GB coupling   the dark energy 
exhibits phantom-like  behavior while for negative values it exhibits 
quintessence-like behavior. Interestingly enough, at early and late times   
$w_{DE}$ tends to the cosmological constant value $-1$ and the deviation 
happens only at intermediate redhsifts, with a peak  at around  $z\approx 2$, a 
behavior that was expected since at early times there are no stars, while at 
asymptotically late times most black holes will have merge. 
Furthermore, for small absolute  
$\tilde{\alpha}$ values the scenario tends to  $\Lambda$CDM paradigm. Finally, 
we investigated the effect of the other model parameters, showing that 
increasing  \( f_{\text{BH}} \) enhances  the deviations from  $\Lambda$CDM 
scenario, while $f_\text{merge}$ has only a minor effect. Nevertheless, 
for  the  whole allowed estimated ranges of the parameters, $w_{DE}$ 
remains within its observational bounds found by Planck Collaboration. 
Finally, note that in the present manuscript we covered all possible 
$\tilde{\alpha}$ cases, and thus all possible $w_{DE}$ behavior, for 
completeness.
 Which subclass is the one favored by nature   can   arise only through 
full confrontation with 
observations.

In summary, by applying the gravity-thermodynamics conjecture with the Wald 
entropy in the case of Einstein-Gauss-Bonnet theory, we obtained an effective, 
topological dark energy sector with interesting cosmological applications. 
Definitely, there are many investigations that need to be done before we 
consider the scenario at hand as a viable candidate for the description of 
nature. One could estimate with higher accuracy the BH merging rate per 
redshift, using also the future accumulating data of LIGO and VIRGO 
observations 
\cite{2018ApJ...864L..19F,Ellis:2020lxl,Barausse:2020mdt,Banks:2021jwj,
Li:2022wxi}.   Additionally, one should use observational data  from Supernova 
type Ia (SNIa), Baryon Acoustic Oscillations 
(BAO), overdensity perturbations $f\sigma 8$ and Hubble rate measurements from 
cosmic chronometers (CC), in order to constrain the involved parameter space.
These studies lie beyond the scope of this manuscript and will be performed in 
future projects.

\subsection*{Acknowledgments} 
The authors would like to acknowledge the  contribution of the LISA 
CosWG, and of   COST 
Actions    CA21136 ``Addressing observational tensions in cosmology with 
systematics and fundamental physics (CosmoVerse)'' and CA23130 ``Bridging high 
and low energies in search of quantum gravity 
(BridgeQG)''.

\bibliography{biblio}

\end{document}